\documentclass[amsmath,jcp,superscriptaddress,twocolumn]{revtex4-1}
\usepackage{color}
\usepackage{amsmath}
\usepackage{amsfonts}
\usepackage{amssymb}
\usepackage{MnSymbol}
\usepackage{graphicx}

\begin{document}
\title{Liouvillian exceptional points of an open driven two-level system}
\author{Nikhil Seshadri}
\affiliation{Harvard University, Cambridge, MA 02138, USA}
\author{Anqi Li}
\affiliation{Department of Chemistry \& Biochemistry, University of California San Diego, La Jolla, CA 92093, USA}
\author{Michael Galperin}
\email{migalperin@ucsd.edu}
\affiliation{Department of Chemistry \& Biochemistry, University of California San Diego, La Jolla, CA 92093, USA}

\begin{abstract}
We study the applicability of the Liouvillian exceptional points (LEPs) approach to nanoscale open
quantum systems. A generic model of the driven two-level system in a thermal environment is analyzed 
within the nonequilibrium Green's function (NEGF) and Bloch quantum master equation (QME) formulations.
We derive the latter starting from the exact NEGF Dyson equations and highlight the
qualitative limitations of the LEP treatment by examining the approximations
employed in its derivation. 
We find that non-Markov character of evolution in open quantum systems does not allow
for the introduction of the concept of exceptional points for a description of their dynamics.
Theoretical analysis is illustrated with numerical simulations.
\end{abstract}

\maketitle

\section{Introduction}\label{intro}
Non-Hermitian quantum mechanics~\cite{moiseyev_non-hermitian_2011} is an accepted way of treating 
open quantum systems which is employed in many fields of theoretical research from optics, opto-mechanics,
and polaritonics, to quantum field theory, molecular physics, and quantum transport.
Complex values of operator spectra in these considerations reflect the non-stationary character of system
states with the balance between gain and loss accounted for by the imaginary parts of eigenvalues.
The most non-trivial physics (such as unidirectional transport, anomalous lasing and absorption, and chiral modes) 
takes place at and in the vicinity of the degeneracies of the complex eigenvalues -- exceptional points (EPs). 

Experimentally, EP behavior has been observed mostly in optics~\cite{miri_exceptional_2019}, in the setting of
a chaotic optical microcavity~\cite{lee_observation_2009},
optical coupled systems with a complex index potential~\cite{ruter_observation_2010},
and photonic lattices~\cite{hahn_observation_2016}.
EP systems were suggested as a platform for development of topological optoelectronics~\cite{doppler_dynamically_2016,ergoktas_topological_2022}.
Recently, observations of EPs in single-spin systems (nitrogen-vacancy centers in diamonds) 
were also reported~\cite{wu_observation_2019}.  
The sensitivity of EP system responses to parameter changes led to
suggestions of employing EP systems as optical~\cite{wiersig_review_2020,mu_realization_2023}
and quantum~\cite{liang_observation_2023} sensors. 
Decoherence enhancement observed in the vicinity of EPs~\cite{zhang_a_2018,naghiloo_quantum_2019} 
opens a way for the exploration of EPs for quantum information processing.
EP physics was also observed in polaritonic systems (exciton-polaritons in semiconductor microcavities)~\cite{gao_observation_2015} and in thermal transport (chiral heat transport)~\cite{xu_non-hermitian_2023}.

The majority of theoretical considerations use effective non-Hermitian Hamiltonians as operators describing
EP physics~\cite{gunther_projective_2007,rotter_non-hermitian_2009,toroker_relation_2009,uzdin_observability_2011,heiss_physics_2012,garmon_analysis_2012,delga_theory_2014,rotter_review_2015,yang_phonon_2020,engelhardt_unusual_2022,ferrier_unveiling_2022,engelhardt_polariton_2023,li_speeding_2023}. These operators are formed by adding complex absorbing potentials (retarded and/or 
advanced projections of self-energies) to Hermitian system Hamiltonians. 
Their degeneracies, the Hamiltonian EPs (HEPs), are the focus of these studies.

Another non-Hermitian operator describing the evolution of open quantum systems is the Liouvillian.
Its degeneracies, Liouvillian EPs (LEPs), were also discussed recently~\cite{pick_robust_2019,minganti_quantum_2019,minganti_hybrid-liouvillian_2020,arkhipov_liouvillian_2020,chimczak_the_2023} . 
Analytical studies comparing HEPs and LEPs conclude that the two types of EPs have 
essentially different properties and that they become equivalent only in the semiclassical limit.
Similar to HEPs, Liouvillian based analysis predicts non-trivial behavior at or in vicinity of LEPs.
For example, LEPs were shown to represent a threshold between diffusive and ballistic motion
in a 1d quantum Lorentz gas~\cite{hashimoto_on_2016,hashimoto_physical_2016}.
Enhancement of decoherence rate~\cite{chen_quantum_2021,khandelwal_signatures_2021,larson_exceptional_2023},
possibility of chiral state transfer~\cite{chen_decoherence-induced_2022},
and optimization of steering towards a predesigned target state~\cite{kumar_optimized_2022}
are predicted in the presence of LEPs. Finally, recent experiment demonstrated enhanced performance
of the single-ion quantum heat engine from the LEPs~\cite{bu_enhancement_2023}.

Recently, we studied applicability of the concept of HEPs in nanoscale open quantum systems~\cite{mukamel_exceptional_2023}.
Utilizing a model of two vibrational modes in a cavity we compared standard nonequilibrium Green's function
(NEGF) with HEP based predictions. We derived the latter from the former and discussed approximations
required to reduce exact NEGF to approximate HEP description. 
In particular, we showed that HEP disregards lesser and greater projections of self-energy
due to intra-system interactions while keeping its retarded projection which makes the HEP 
treatment inconsistent and may lead to qualitative failures. Another limiting factor of the HEP
approach is its Markov character.

Here, we present analysis of LEP based considerations starting from exact NEGF treatment and
exploring approximations necessary to reduce the latter to the approximate LEP description.
The two most basic and widely employed models for LEP analysis are  
the driven two-level system (TLS)~\cite{am-shallem_exceptional_2015,hatano_exceptional_2019,perinajr_quantum_2022}
and oscillator~\cite{tay_liouvillian_2023} in a generic environment.
We use the TLS as a model for comparison between NEGF and LEP methods. 
Similar to our findings in Ref.~\onlinecite{mukamel_exceptional_2023},
LEP is also limited by its Markov character. Nevertheless, contrary to the HEP, the Liouvillian based treatment
disregards the retarded projection of the self-energies while keeping their lesser and greater projections.
Some limitations in the applicability of LEP methods to nanoscale open quantum systems are
illustrated with simulations comparing the NEGF and Bloch quantum master equation (QME) 
results for driven TLS in a thermal environment.

In Section~\ref{TLS} we introduce the model and
present its NEGF treatment. We then utilize NEGF as a starting point for derivation of
the Bloch QME and its generalization which accounts for dissipation and
discuss approximations necessary to reduce exact NEGF treatment to approximate
Redfield/Lindblad QME. 
Section~\ref{numres} compares results of numerical simulations performed within the NEGF formulation and within 
the two types of the Bloch QME formulations. Conclusions are drawn in Section~\ref{conclude}.

\section{Driven TLS in a thermal environment}\label{TLS}
\subsection{Model}
We consider a two-level system that is driven by external classical field $E(t)$ and dissipated by a thermal bath.
The latter is continuum of Bose modes $\{\alpha\}$. The Hamiltonian of this model is
\begin{equation}
\hat H(t) = \hat H^S(t) + \hat H^B + \hat V^{SB}
\end{equation}
where $\hat H^S(t)$ and $\hat H^B$ describe decoupled system and bath, respectively.
$\hat V^{SB}$ is the system-bath coupling. Explicit expressions for each of the terms are given by 
\begin{equation}
\begin{split}
\hat H^S(t) &= \sum_{i=1,2}\varepsilon_i\hat d_i^\dagger\hat d_i
-\mu E(t)\left(\hat d_1^\dagger\hat d_2+\hat d_2^\dagger\hat d_1\right)
\\
\hat H^B &= \sum_\alpha\omega_\alpha\hat b_\alpha^\dagger\hat b_\alpha
\\
\hat V^{SB} &= \sum_{i,j=1,2}\sum_\alpha\left(
V_{ij,\alpha}\left[\hat d_i^\dagger\hat d_j\right]^\dagger\hat b_\alpha +
V_{\alpha,ij}\,\hat b^\dagger_\alpha\left[\hat d_i^\dagger\hat d_j\right]
\right)
\end{split}
\end{equation}
Here, $\hat d_i^\dagger$ ($\hat d_i$) and $\hat b_\alpha^\dagger$ ($\hat b_\alpha$)
creates (annihilates) an electron in level $i$ and an excitation in mode $\alpha$, respectively.
$\mu$ is the transition dipole moment. The driving function is taken to be harmonic 
\begin{equation}
\label{Et}
E(t)=E_0\cos(\omega_0 t)
\end{equation}
In the following analysis, we assume $\varepsilon_1<\varepsilon_2$
and consider coupling to the thermal bath in the rotating-wave approximation (RWA);
that is, $V_{21,\alpha}=V_{\alpha,21}=0$.
We note that the RWA is central for the derivation of the Bloch QME. 

\begin{figure}[htbp]
\centering\includegraphics[width=0.8\linewidth]{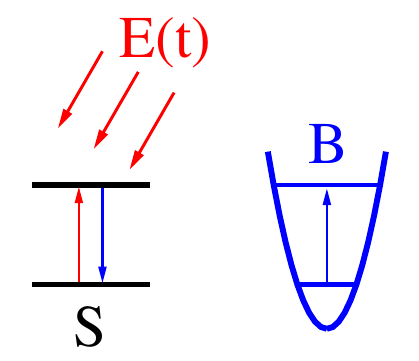}
\caption{\label{fig1}
 Sketch of a model for optically driven two-level system (S) in thermal environment (B).
}
\end{figure}


\subsection{NEGF formulation}
Within the NEGF formulation, the central quantity of interest is the single-particle Green's function of the system
defined on the Keldysh contour
\begin{equation}
\label{Gij_def}
 G_{ij}(\tau_1,\tau_2) \equiv -i\langle T_c\, \hat d_i(\tau_1)\,\hat d_j^\dagger(\tau_2)\rangle.
\end{equation}
Here, $T_c$ is the contour ordering operator, $\tau_{1,2}$ are the contour variables,
and the creation (annihilation) operator $\hat d_j^\dagger(\tau_2)$ ($\hat d_i(\tau_1)$) is in the Heisenberg picture.
Knowledge of $G_{ij}(\tau_1,\tau_2)$ allows for the calculation of characteristics of the system and its responses to
external perturbations. In particular, in the single-electron subspace of the problem, the system density matrix 
is given by the lesser projection of the Green's function (\ref{Gij_def})
taken at equal times
\begin{equation}
\label{rhoij}
 \rho_{ij}(t) = -i\, G^{<}_{ij}(t,t)
\end{equation}
This relation is central for comparison between the NEGF and Bloch quantum master equation (QME) results.

The dynamics of the system is described by the Dyson equation for the Green's function (\ref{Gij_def}) 
\begin{align}
\label{Dyson}
&i\frac{\partial}{\partial\tau_1} G_{ij}(\tau_1,\tau_2) = \delta_{i,j}\,\delta(\tau_1,\tau_2)
\\ &
+\sum_{n=1,2}H_{in}^S(t_1)\, G_{nj}(\tau_1,\tau_2) 
+\int_c d\tau\, \Sigma_{in}(\tau_1,\tau)\, G_{nj}(\tau,\tau_2)
 \nonumber
\end{align}
where $t_1$ is physical time corresponding to contour variable $\tau_1$
and $\Sigma(\tau_1,\tau)$ is the self-energy due to the coupling of the system to the bath.
While the exact expression for the latter is not accessible due to the many-body character of the system-bath coupling 
$\hat V^{SB}$, an appropriate level of theory for future comparison with the Bloch QME can be achieved by a
second order diagrammatic expansion. Within this (Hartree-Fock) approximation, the expression for the self-energy is
(see Appendix~\ref{appA} for derivation)
\begin{align}
\label{Sigma}
&\Sigma_{ij}(\tau_1,\tau_2) = 
\delta(\tau_1,\tau_2)\, \sum_{n_1,n_2}\int_c d\tau\,\rho_{n_1n_2}(t)
\nonumber \\ &\qquad\qquad\,\,\,\times
\bigg[\Pi_{ji,n_2n_1}(\tau_1,\tau)+\Pi_{n_1n_2,ij}(\tau,\tau_1)\bigg]
\\ &+
i\sum_{n_1,n_2}\bigg[\Pi_{n_1i,n_2j}(\tau_1,\tau_2)+\Pi_{jn_2,in_1}(\tau_2,\tau_1)\bigg]
G_{n_1n_2}(\tau_1,\tau_2)
\nonumber
\end{align}
Here,
\begin{equation}
\label{Pi_def}
\Pi_{n_1n_2,n_3n_4}(\tau_1,\tau_2) \equiv \sum_\alpha V_{n_1n_2,\alpha}\, F^{(0)}_\alpha(\tau_1,\tau_2)\,
V_{\alpha,n_3n_4}
\end{equation}
is the thermal bath-induced effective interaction between transitions $n_1n_2$ and $n_3n_4$, and
\begin{equation}
\label{F_def}
 F^{(0)}_\alpha(\tau_1,\tau_2) \equiv -i\left\langle T_c\, \hat b_\alpha(\tau_1)\,
 \hat b_\alpha^\dagger(\tau_2)\right\rangle_0
\end{equation}
is the Green's function of free phonon mode $\alpha$ in the bath.


\subsection{Bloch QME}
The derivation of an approximate Redfield/Lindblad QME starts from
the exact equation-of-motion (EOM) for the density matrix given by (\ref{rhoij}), which is derived within the NEGF formulation.
The EOM is (see Appendix~\ref{appB} for derivation)
\begin{widetext}
\begin{align}
\label{rho_EOM}
\frac{d}{dt}\rho_{ij}(t) &= i\,\omega_{ji}\,\rho_{ij}(t) 
-i\,\mu\, E(t)\left[ \rho_{i\bar j}(t)-\rho_{\bar i j}(t) \right]
\\ &+ \sum_{n,n_1,n_2}\int_{-\infty}^{t} dt'\,\bigg[
\Pi^{>}_{ni,n_1n_2}(t-t')\, G^{(2)\, <}_{n_1n_2,nj}(t',t) - \Pi^{<}_{ni,n_1n_2}(t-t')\, G^{(2)\, >}_{n_1n_2,nj}(t',t)
\nonumber \\ &\qquad\qquad\qquad\,\,\,\,\,
- \Pi^{>}_{jn,n_1n_2}(t-t')\, G^{(2)\, <}_{n_1n_2,in}(t',t) + \Pi^{<}_{jn,n_1n_2}(t-t')\, G^{(2)\, >}_{n_1n_2,in}(t',t)
\nonumber \\ &\qquad\qquad\qquad\,\,\,\,\,
+ G^{(2)\, <}_{ni,n_1n_2}(t,t')\,\Pi^{>}_{n_1n_2,nj}(t'-t) - G^{(2)\, >}_{ni,n_1n_2}(t,t')\,\Pi^{<}_{n_1n_2,nj}(t'-t)
\nonumber \\ &\qquad\qquad\qquad\,\,\,\,\,
- G^{(2)\, <}_{jn,n_1n_2}(t,t')\, \Pi^{>}_{n_1n_2,in}(t'-t)
+ G^{(2)\, >}_{jn,n_1n_2}(t,t')\,\Pi^{<}_{n_1n_2,in}(t'-t) 
\bigg]
\nonumber
\end{align}
where $\bar i,\bar j=2 (1)$ for $i, j=1 (2)$, $\omega_{ji}\equiv\varepsilon_j-\varepsilon_i$, and
\begin{equation}
\label{G2_def}
G^{(2)}_{n_1n_2,n_3n_4}(\tau_1,\tau_2) \equiv -i\langle T_c\, 
\left[\hat d_{n_1}^\dagger\hat d_{n_2}\right](\tau_1)\,
\left[\hat d_{n_3}^\dagger\hat d_{n_4}\right]^\dagger(\tau_2)\rangle
\end{equation}
is the two-particle Green's function.
Reducing the exact EOM (\ref{rho_EOM}) to the Redfield/Lindblad QME requires 
approximating its right side with a Markov dynamics.
The Redfield/Lindblad QME can be obtained from the Green's function Dyson equation by employing 
a Kadanoff-Baym-like ansatz~\cite{haug_quantum_2008} 
\begin{equation}
\label{KB}
 G^{(2)\,\gtrless}_{n_1n_2,n_3n_4}(t_1,t_2) \approx
 \theta(t_1-t_2)\, e^{-i\omega_{21}(t_1-t_2)}\, G^{(2)\,\gtrless}_{n_1n_2,n_3n_4}(t_2,t_2)
 +\theta(t_2-t_1)\, e^{-i\omega_{43}(t_1-t_2)}\, G^{(2)\,\gtrless}_{n_1n_2,n_3n_4}(t_1,t_1)
\end{equation}
where $\theta(\ldots)$ is the Heaviside step function.
Employing this ansatz leads to the Bloch equations (see Appendix~\ref{appC} for derivation)
\begin{equation}
\label{rho_Bloch}
\frac{d}{dt}\begin{pmatrix} \rho_{11}(t) \\ \rho_{22}(t) \\ \rho_{12}(t) \\ \rho_{21}(t) \end{pmatrix} =
-i \begin{bmatrix}
-i W_{2\leftarrow 1} & iW_{1\leftarrow 2} & \mu E(t) & -\mu E(t) \\
iW_{2\leftarrow 1} & -iW_{1\leftarrow 2} & -\mu E(t) & \mu E(t) \\
\mu E(t) & -\mu E(t) & -\omega_{21}-iW_d & 0 \\
-\mu E(t) & \mu E(t) & 0 & \omega_{21} -iW_d
\end{bmatrix}
\begin{pmatrix}
\rho_{11}(t) \\ \rho_{22}(t) \\ \rho_{12}(t) \\ \rho_{21}(t)
\end{pmatrix}
\end{equation}
\end{widetext}
Here,
\begin{equation}
\label{W12}
\begin{split}
 W_{2\leftarrow 1} &\equiv \Gamma_{12,12}(\omega_{21})\, N(\omega_{21}),
 \\
W_{1\leftarrow 2} &\equiv \Gamma_{12,12}(\omega_{21})\, \left[1+N(\omega_{21})\right],
\end{split}
\end{equation}
are the population transfer rates, and
\begin{equation}
\label{Wd}
\begin{split}
& W_d \equiv \frac{W_{2\leftarrow 1}+W_{1\leftarrow 2}}{2}
\\ &\quad\,\,\,
 + \frac{\Gamma_{11,11}(0)+\Gamma_{22,22}(0)}{2}\left[1+2\, N(0)\right]
\end{split}
\end{equation}
is the dephasing rate. In Eqs.~(\ref{W12})-(\ref{Wd})
\begin{equation}
 \Gamma_{n_1n_2,n_3n_4}(\omega)\equiv 2\pi\sum_{\alpha} V_{n_1n_2,\alpha}\, V_{\alpha,n_3n_4}\,\delta(\omega-\omega_\alpha)
\end{equation}
is the dissipation matrix.

Finally, by using (\ref{Et}) with 
$\omega_0-\omega_{21}\equiv \Delta\ll\omega_0$, going into rotating frame of the field
\begin{equation}
\tilde\rho_{12}(t) \equiv e^{-i\omega_0 t}\,\rho_{12}(t),
\end{equation}
and introducing the spin operators
\begin{equation}
\begin{split}
\tilde S_x(t) &\equiv \tilde\rho_{21}(t)+\tilde\rho_{12}(t)
\\
\tilde S_y(t) &\equiv i\left[\tilde\rho_{21}(t)-\tilde\rho_{12}(t)\right]
\\
S_z(t) &\equiv\rho_{22}(t)-\rho_{11}(t)
\end{split}
\end{equation}
one can employ the rotating wave approximation (RWA) to express the Bloch QME (\ref{rho_Bloch})
as an EOM for the spin operator
\begin{equation}
\label{Bloch_spin}
\begin{split}
\frac{d}{dt}
\begin{pmatrix}
\tilde S_x(t) \\ \tilde S_y(t) \\ S_z(t)
\end{pmatrix} &=
\begin{bmatrix}
-\frac{1}{T_2} & \Delta & 0 \\
-\Delta & -\frac{1}{T_2} & \mu E_0 \\
0 & -\mu E_0 & -\frac{1}{T_1}
\end{bmatrix}
\begin{pmatrix}
\tilde S_x(t) \\ \tilde S_y(t) \\ S_z(t)
\end{pmatrix} 
+
\begin{pmatrix}
0 \\ 0 \\ \frac{S_z^{0}}{T_1}
\end{pmatrix}
\end{split}
\end{equation}
Here,
\begin{equation}
\begin{split}
&\frac{1}{T_1}\equiv W_{2\leftarrow 1}+W_{1\leftarrow 2},\qquad
\frac{1}{T_2}\equiv W_d,
\\ &
S_z^{0}\equiv\frac{W_{2\leftarrow 1}-W_{1\leftarrow 2}}{W_{2\leftarrow 1}+W_{1\leftarrow 2}}
\end{split}
\end{equation}


\subsection{Generalized Bloch QMEs}
While deriving the Bloch QME (\ref{rho_Bloch}) from the exact EOM (\ref{rho_EOM}),
one looses proper non-Markov evolution and disregards dissipation.
Note that while the former is common for Hamiltonian and Liouvillian EP formulations~\cite{mukamel_exceptional_2023},
the latter is specific to Liouvillian EPs. Indeed, the Hamiltonian EP formulation disregards 
the lesser/greater projections of the self-energy, while the ansatz (\ref{KB}) misses the retarded projection, 
however, the generalized Kadanoff-Baym ansatz (GKBA) in the NEGF literature~\cite{lipavsk_y_generalized_1986,haug_quantum_2008}
does preserve information about dissipation. To construct the Liouville space analog we follow 
procedure originally introduced in Ref.~\onlinecite{esposito_transport_2009}. This leads to
(see Appendix~\ref{appD} for derivation)
\begin{equation}
\label{GKB}
\begin{split}
 & G^{(2)\,\gtrless}_{n_1n_2,n_3n_4}(t_1,t_2) \approx
 \\ &\qquad
 i\sum_{e,f}\bigg[\mathcal{G}^r_{n_1n_2,ef}(t_1-t_2)\, G^{(2)\,\gtrless}_{ef,n_3n_4}(t_2,t_2)
 \\ &\qquad\quad\,\,
 -G^{(2)\,\gtrless}_{n_1n_2,ef}(t_1,t_1)\,\mathcal{G}^a_{ef,n_3n_4}(t_1-t_2)\bigg]
\end{split}
\end{equation}
where
\begin{equation}
\label{Gra_Liouville}
\begin{split}
\mathcal{G}^r_{n_1n_2,n_3n_4}(t_1,t_2) &\equiv -i\theta(t_1-t_2) \llangle n_2n_1\rvert\, \mathcal{U}_{eff}(t_1,t_2)\,
\lvert n_4n_3\rrangle
\\
\mathcal{G}^a_{n_1n_2,n_3n_4}(t_1,t_2) &\equiv +i\theta(t_2-t_1) \llangle n_2n_1\rvert\, \mathcal{U}^\dagger_{eff}(t_2,t_1)\,
\lvert n_4n_3\rrangle
\end{split}
\end{equation}
are the retarded and advanced Green's functions in Liouville space, respectively, and
\begin{equation}
\mathcal{U}_{eff}(t_1,t_2) \equiv  T\,\exp\left[-i\int_{t_2}^{t_1}dt\, \mathcal{L}_{eff}(t)\right]
\end{equation}
is the Liouville space effective evolution operator, where $\mathcal{L}_{eff}(t)$ defines time evolution in 
the system subspace of the problem.
We note that Eq.(\ref{GKB}) is an approximation. The approximation is introduced by 
employing projection operator (\ref{proj}) in exact expressions (\ref{G2_Liouville})
which makes (\ref{GKB}) a second order contribution in infinite diagrammatic expansion of the 
coupled system-bath evolution in strength of the system-bath coupling.

We employ parts of the Redfield/Lindblad Liouvillian
matrix in the right side of Eq.(\ref{rho_Bloch}), as the system evolution generator.
In particular, retaining only free evolution (i.e disregarding driving $\mu E(t)$ and dissipation
$W_{2\leftarrow 1}$, $W_{1\leftarrow 2}$, and $W_d$) reduces (\ref{GKB}) to (\ref{KB}).

Keeping the dissipation, using (\ref{GKB}) in (\ref{rho_EOM}), and assuming the Born-Markov approximation 
leads to a generalized version of the Bloch QME, which retains the same form (\ref{rho_Bloch}),
although with renormalized ($\bar W$) dissipation rates
\begin{align}
\label{barW}
&\bar W_{2\leftarrow 1} \equiv -i\int\frac{d\omega}{2\pi}\, \Pi^{<}_{12,12}(\omega)\,
\mbox{Im}\left[\frac{1}{\omega-\omega_{21}+iW_d}\right]
\nonumber \\
&\bar W_{1\leftarrow 2} \equiv -i\int\frac{d\omega}{2\pi}\, \Pi^{>}_{12,12}(\omega)\,
\mbox{Im}\left[\frac{1}{\omega-\omega_{21}+iW_d}\right]
\nonumber \\
&\bar W_d \equiv -\int\frac{d\omega}{2\pi}\,\bigg[
\frac{\Pi^{<}_{11,11}(\omega)+\Pi^{>}_{22,22}(\omega)}{\omega+iW_d}
\\ &
- \frac{\Pi^{>}_{11,11}(\omega)+\Pi^{<}_{22,22}(\omega)}{\omega-iW_d}
\nonumber \\ &
+\frac{V^R_{11,1}\, \left[ V^{R}\right]^{-1}_{1,11}\, \Pi^{>}_{12,12}(\omega)+V^R_{22,1}\, \left[ V^{R}\right]^{-1}_{1,22}\,\Pi^{<}_{12,12}(\omega)}{\omega-\omega_{21}+i\delta}
\nonumber \\ &
+\frac{V^R_{11,2}\, \left[ V^{R}\right]^{-1}_{2,11}\, \Pi^{>}_{12,12}(\omega)+V^R_{22,2}\, \left[ V^{R}\right]^{-1}_{2,22}\,\Pi^{<}_{12,12}(\omega)}{\omega-\omega_{21}+i\left(W_{2\leftarrow 1}+W_{1\leftarrow 2}\right)}
\bigg]
\nonumber
\end{align}
Here, $\mathbf{V}^R$ is the right eigenvector of the Liouvillian matrix.
Note that while also keeping driving terms in the effective evolution
is possible, we will not pursue this direction because the accepted approach regarding the derivation 
of the standard Bloch QME requires one  to disregard the driving term when deriving dissipators of the Liouvillian.
Note also, that using the Liouville space generalized Kadanoff-Baym ansatz on the Keldysh 
anti-contour~\cite{esposito_self-consistent_2010} would lead to the same form of the
generalized Bloch equation.

Finally, one can choose to solve the time-nonlocal (non-Markov) version of the QME.
Using (\ref{GKB}) in (\ref{rho_EOM}) without the Born-Markov assumption leads to
\begin{equation}
\label{nonMarkov}
\begin{split}
&\frac{d}{dt}\rho_{11}(t) =\mu\, E(t)\, 2\,\mbox{Im}\left[\rho_{12}(t)\right]
\\ &
+2\,\mbox{Im}\int_{-\infty}^t dt'\,\bigg(
\Pi^{<}_{12,12}(t-t')\,\rho_{11}(t')
\\ &\qquad
-\Pi^{>}_{12,12}(t-t')\left[1-\rho_{11}(t')\right]
e^{(i\omega_{21}-W_d)(t-t')} \bigg)
\\
& \frac{d}{dt}\rho_{12}(t) = i\omega_{12}\rho_{12}(t)-i\mu\, E(t)\left[2\,\rho_{11}(t)-1\right]
\\ &
-i\int_{-\infty}^t dt' \bigg(
\left[\Pi_{11,11}^{>}(t-t')+\Pi_{22,22}^{<}(t-t')\right.
\\ & \quad
\left. +\Pi_{11,11}^{<}(t'-t)+\Pi_{22,22}^{>}(t'-t)\right]
e^{(i\omega_{21}-W_d)(t-t')}
\\ &
+\sum_{i=1,2}\left[ V^R_{11,i}\, e^{-i\lambda_i(t-t')}\,\left[V^R\right]^{-1}_{i,11}\,\Pi^{>}_{12,12}(t'-t) \right.
\\ & \quad
\left. + V^R_{22,i}\, e^{-i\lambda_i(t-t')}\,\left[V^R\right]^{-1}_{i,22}\,\Pi^{<}_{12,12}(t'-t) \right]
\bigg)\rho_{12}(t')
\end{split}
\end{equation}
Here, $\lambda_i$ are the eigenvalues of the Liouvillian matrix.
We note that non-Markov version of the Bloch QME, Eq.(\ref{nonMarkov}),
accounts for broadening of system states induced by their hybridization with the bath
which is completely missed by the standard Bloch QME, Eq.(\ref{rho_Bloch}).
At the same time,  this result is still an approximation 
(it is only second order in infinite hybridization expansion).
That is, while for moderate coupling strengths Eq.(\ref{nonMarkov}) 
can produce relatively accurate results,
for significant system-bath coupling strengths the approximation may fail.

Below we use the Bloch equation (\ref{rho_Bloch}) and its generalizations (\ref{barW}) and (\ref{nonMarkov})
to discuss the concept of exceptional points for a Liouville operator.
Following Ref.~\onlinecite{am-shallem_exceptional_2015}, we evaluate the time dependence of
the $z$-projection of the spin operator $S_z(t)$ and use it in eigeinmode analysis
\begin{equation}
\label{eigenmode}
 S_z(t) = \sum_k d_k\, e^{-i\omega_k t}
\end{equation}
Degeneracies of the complex eigenmodes $\omega_k$ represent LEPs.
As discussed in Ref.~\onlinecite{fuchs_harmonic_2014}, the latter can be approximately found from the points of
divergence of the absolute values of the coefficients $\lvert d_k\rvert$, although extended analysis is needed
for further characterization. We will use the parameters found for LEPs 
in Ref.~\onlinecite{am-shallem_exceptional_2015} as a starting point for our consideration.


\begin{figure}[b]
\centering\includegraphics[width=0.8\linewidth]{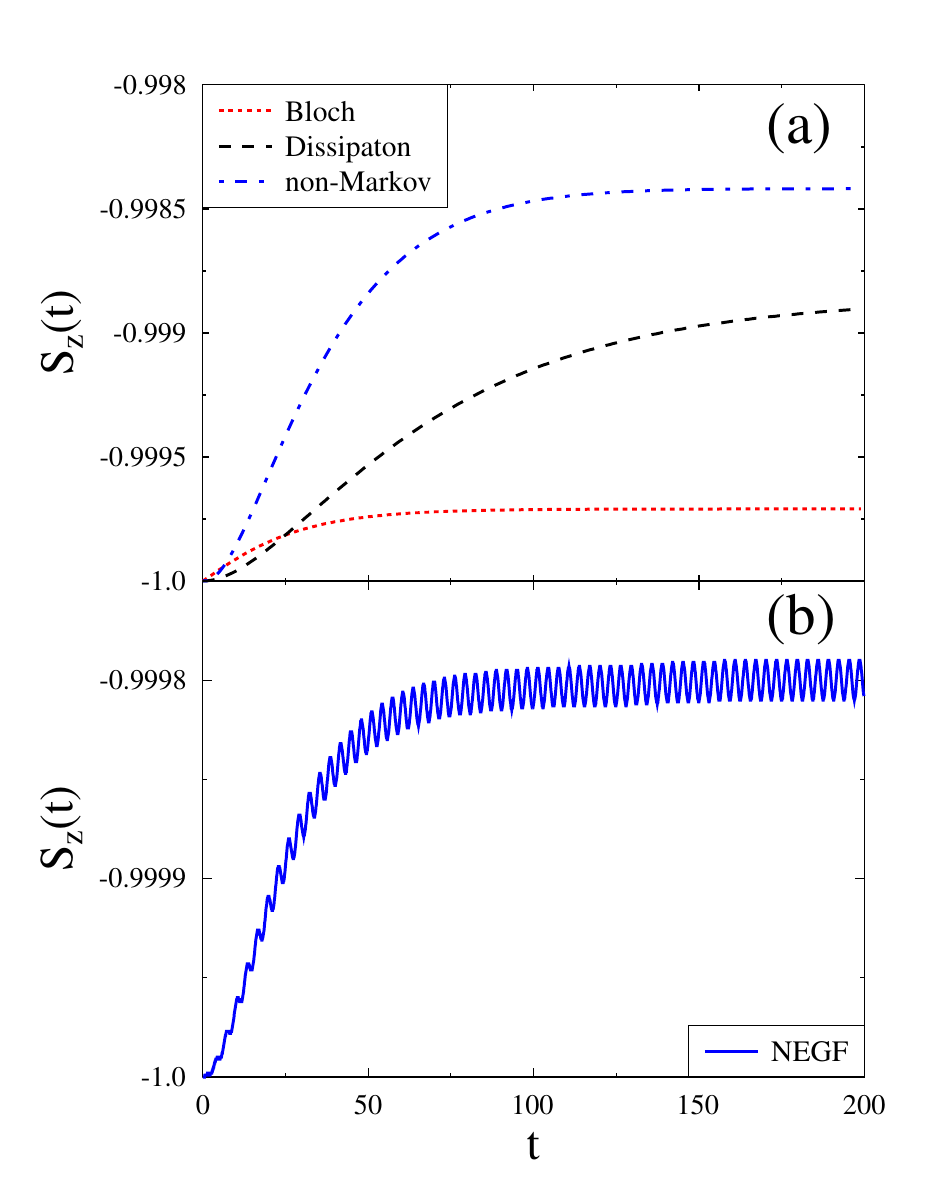}
\caption{\label{fig2}
 Time dependence of the $z$-component of spin operator, $S_z(t)$.
 Panel (a) presents the results of simulations employing 
 the standard Bloch QME, Eq.(\ref{Bloch_spin}) (dotted line, red); 
 Bloch QME with dissipation included, Eq.~(\ref{barW}) (dashed line, black);
 and non-Markov Bloch QME with dissipation included, Eq.(\ref{nonMarkov}) (dash-dotted line, blue).
 Panel (b) shows the results of the NEGF Hartree-Fock simulations, Eqs.~(\ref{rhoij})-(\ref{Sigma}).
}
\end{figure}

\section{Numerical results}\label{numres}
We now evaluate $S_z(t)$ within different methodologies and use the results of simulations
to obtain exceptional points for the Liouville operator. 

Unless stated otherwise, the parameters of the simulations are the following. The energy levels of the system are 
$\varepsilon_1=0$ and $\varepsilon_2=1$, the laser detuning $\Delta\equiv\omega_0-\omega_{21}=0.00102$,
and the coupling to driving field $\mu\, E_0=0.001$. For simplicity we take $V_{\alpha,11}=V_{\alpha,22}=0$,
so that the dephasing rates are $1/T_1=2/T_2=0.1$. The temperature of the bath is assumed to be zero. 
Simulations were performed on a time grid of $200$ points with step $1$. We confirmed that simulations
on a grid of $2000$ points with step $0.1$ yield similar results.

For non-Markov simulations we employ the bath spectral function
\begin{equation}
 J(\omega)\equiv \left(\frac{\omega}{\omega_{21}}\right)^2\,\exp\left[2\left(1-\frac{\omega}{\omega_{21}}\right)\right]
\end{equation}
and the bath dephasing rate is defined as
\begin{equation}
\Gamma_{12,12}(\omega)\equiv\Gamma_{12,12}(\omega)\, J(\omega) = J(\omega)/T_1
\end{equation}
Fast Fourier transform is performed on a grid of 10001 points and utilizes the FFTW library~\cite{FFTW}.
The NEGF simulations are performed by employing the procedure first introduced in Ref.~\cite{stan_time_2009}.

Figure~\ref{fig2} shows time dependence of the $z$-projection of the spin operator
after employing the various approaches. Note that the
differences in shapes of the curves $S_z(t)$ reflect differences in underlying eigenmode compositions.
Note that the differences in long time value of the projection are due to renormalization of the dissipation
parameters (\ref{barW}) and thus are of secondary importance. 
Comparing QME and NEGF results (panels (a) and (b), respectively) we note 
oscillating behavior of the NEGF stationary state and difference in magnitude of the signal.
The reason for the discrepancy are assumptions made when deriving the Bloch QME and its 
generalizations: 1.~the rotating wave approximation in external driving and
2.~neglecting effect of the driving term on dissipator super-operator 
(i.e. dissipator is derived as if there is no driving). 
Within the NEGF, the driving term is taken into account exactly.

We now turn to the exceptional points analysis of the resulting time series.
 Because we use the parameters of Ref.~\onlinecite{am-shallem_exceptional_2015}
 we know that our standard Bloch QME simulations are performed in vicinity of LEP of second order.
 Thus, the divergence of coefficient $\lvert d_k\rvert$ indicates the presence of an exceptional point.
 Instead of the harmonic inversion analysis employed in Ref.~\cite{am-shallem_exceptional_2015}
 for eigenmode decomposition, we use its filter diagonalization variant~\cite{mandelshtam_harmonic_1997}.
For the parameters of the simulations, the latter method appears to be more stable.  

\begin{figure}[htbp]
\centering\includegraphics[width=0.8\linewidth]{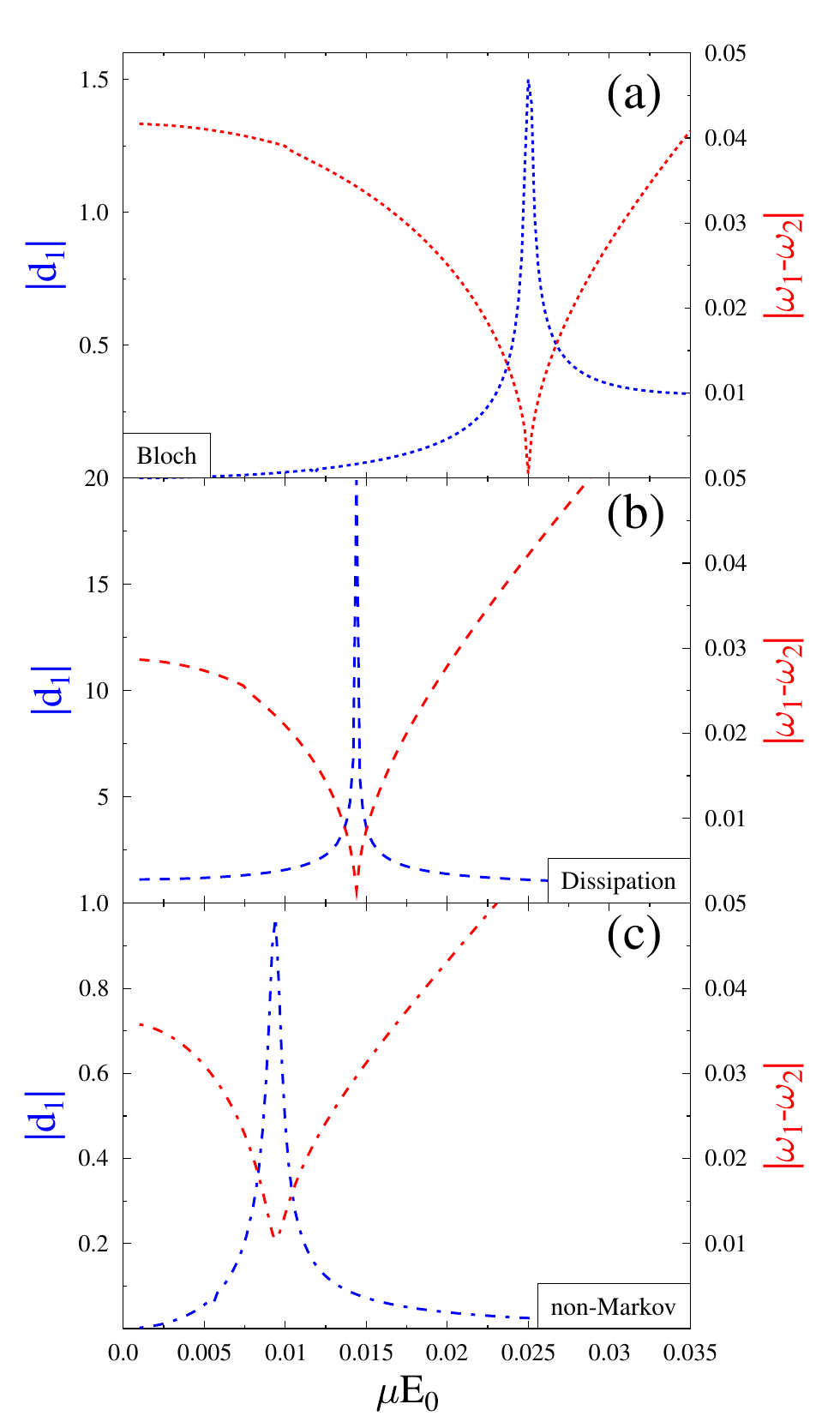}
\caption{\label{fig3}
 Eigenmode analysis for the $z$-component of spin operator $S_z(t)$, Eq.(\ref{eigenmode}).
 Shown are absolute values of the coefficient of expansion $\lvert d_1\rvert$ (blue line, left axis)
 and difference between eigenmodes $\lvert\omega_1-\omega_2\rvert$ (red line, right axis) 
 for results of simulations performed within
 (a) standard Bloch QME, (b) generalized Bloch QME with dissipation included, and
 (c) non-Markov generalized Bloch QME with dissipation included.  
 }
\end{figure}

Figure~\ref{fig3} presents the eigenmode analysis for the time series $S_z(t)$ obtained within different Bloch QME
schemes. The divergence of the expansion coefficient $\lvert d_1\rvert$ and the disappearance of the eigenmode 
difference in the analysis of the standard Bloch QME results presented in panel (a)
indicate the presence of a second order LEP at $\mu\, E_0\sim 0.025$. 
Panel (b) shows similar analysis for generalized Bloch QME with included dissipation.
Similar to the standard Bloch QME three eigenmodes are present in the region away from the LEP.
QME rates renormalization, Eq.(\ref{barW}), leads to a shift of the position of the LEP,
which now occurs at $\mu\, E_0\sim 0.015$.
Result of analysis for the non-Markov QME is shown in panel (c).
We find four different eigenmodes in this case. Careful comparison with Markov consideration
of panel (b) shows absence of exceptional points: one can see that difference in eigenmodes
does not disappear. 

\begin{figure}[t]
\centering\includegraphics[width=0.8\linewidth]{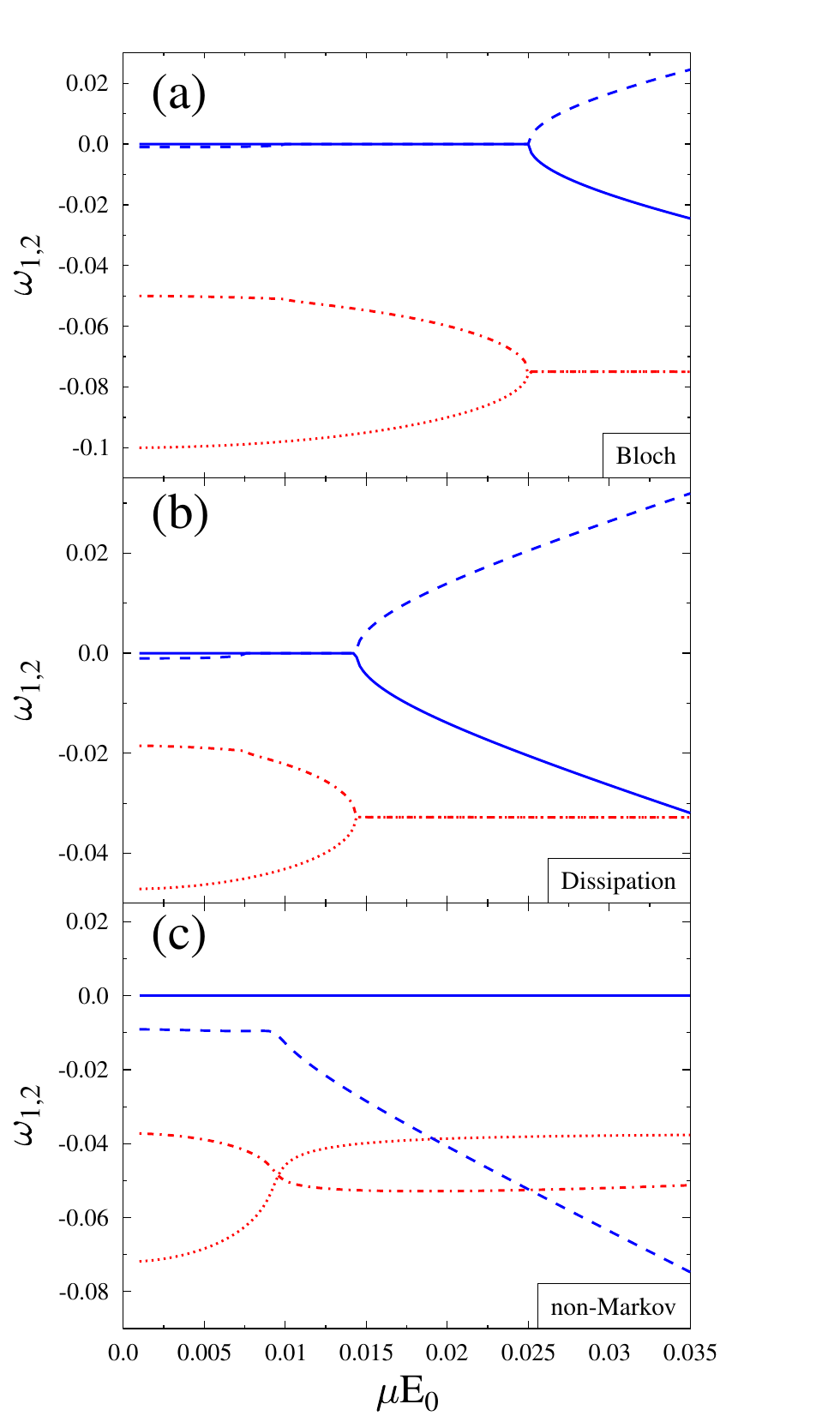}
\caption{\label{fig4}
 Eigenmodes $\omega_1$ and $\omega_2$ vs. coupling to driving field $\mu\, E_0$.
 Shown are real (blue solid and blue dashed lines) and imaginary (red dashed and red dash-dotted lines)
 parts of the eigenmodes for $S_z(t)$ simulations performed within
 (a) standard Bloch QME, (b) generalized Bloch QME with dissipation included, and
 (c) non-Markov generalized Bloch QME with dissipation included.  
 }
\end{figure}

Figure~\ref{fig4} shows two eigenmodes which become degenerate at exceptional point.
One sees that for the standard (panel a) and the generalized (panel b) Markov QME  weak coupling
to the driving field (below the LEP) corresponds to situation where real parts of the eigenmodes
coincide while imaginary parts are different. Stronger couplings (above the LEP)
correspond to zero difference in imaginary parts and different real parts. 
Note that similar behavior at LEP yields transition between diffusive and ballistic motion~\cite{hashimoto_on_2016,hashimoto_physical_2016}
and enhancement of decoherence rate~\cite{chen_quantum_2021,khandelwal_signatures_2021,larson_exceptional_2023}.
Behavior of eigenmodes for results obtained within non-Markov QME (panel c) 
is more complicated. No degeneracy is observed between the modes.
Similarly, eigenmode analysis for the NEGF results yields a large number of modes ($\sim 50$)
with no LEPs present.

The absence of exceptional points in the results of non-Markov evolution is expected 
because the EOM for $S_z(t)$ is not generated by the time-independent Liouvillian anymore. 
One can understand the absence of the LEPs in this case from a purely mathematical perspective.
Indeed, even if one starts from a time-dependent characteristic for a LEP
(for example, for LEP2 one expects to have $\rho(t)\sim (d_1+d_2t) e^{-i\lambda t}$)
the first step of time evolution will annihilate the LEP time dependence due
to convolution of the density operator with the time-dependent function, Eq.(\ref{nonMarkov}).
Indeed, taking the integral in Eq.(\ref{nonMarkov}) with memory kernel which depends on time in 
a complicated way does not preserve original form of $\rho(t)$.
This can be easily seen by expanding the kernel in Fourier series and performing time integration.

Failure of the concept of the Liouvillian exceptional point for non-Markov evolution is 
even more obvious when analyzing the more rigorous NEGF formulation, 
Eqs.~(\ref{Dyson})-(\ref{Pi_def}).
Indeed, in the right side of the Dyson equation one has product of two Green's functions:
one from Eq.(\ref{Dyson}), the other from Eq.(\ref{Sigma}). 
In principle, one could start from Eqs.~(\ref{Dyson})-(\ref{Pi_def}) and apply 
the generalized Kadanoff-Baym ansatz to these expressions. 
This would yield an analog of QME which differs from (but is more accurate than) the Bloch QME. 
Such equation would contain $\rho^2$ in its right side which obviously indicates that 
the form $\rho(t)=(d_1+d_2t)e^{-i\lambda t}$ does not survive in non-Markov formulation. 

We note that the central parameters for the accuracy of Markov approximation are 
the characteristic times of bath $t_c^B$
and system $t_c^S$ dynamics: Markov approximation is accurate when $t_c^B\ll t_c^S$.
For the model, $t_c^B$ is defined by the bandwidth $W_B$, temperature $T$, and structure of 
the bath spectral function $J(\omega)$: 
$t_c^B\sim \hbar/W, \hbar/k_BT$.  The characteristic time of system dynamics is defined
by the intra-system energy parameters (inter-level separation $\omega_{21}$, driving frequency $\omega_0$,
detuning $\Delta$) and by dissipation rate due to coupling to the bath (e.g., $\Gamma_{12,12}$):
$t_c^S\sim 2\pi/\omega_{21}, 2\pi/\omega_0, 2\pi/\Delta, \hbar/\Gamma_{12,12}$.

\section{Conclusion}\label{conclude}
We discuss the concept of Liouvillian exceptional points (LEPs) used in the description of dynamics 
of open quantum systems. The discussion is focused on a model of driven two-level system
coupled to a thermal bath. Starting with exact NEGF formulation of the problem and implementing
set of approximations we derive standard Bloch QME and its generalizations. 
The latter include dissipation (retarded self-energy contribution). One of the generalizations
is non-Markov. 

We compare this approach with our recent publication~\cite{mukamel_exceptional_2023} 
where similar analysis for the Hamiltonian exceptional points (HEPs) was carried out.
We note that both HEP and LEP approximations rely on Markov description of the system
evolution. In terms of neglected self-energies, standard HEP and LEP considerations
are complementary: while HEP disregards lesser and greater projections of self-energy,
standard LEP misses its retarded projection (dissipation).

By performing simulations for parameters previously shown to provide exceptional points~\cite{am-shallem_exceptional_2015} we find that generalized Bloch QME which includes information 
about dissipation and treats evolution as Markov process is capable to provide LEPs 
although for adjusted parameters. The non-Markov character of evolution does not
permit introduction of the concept of LEPs.
In particular, neither the non-Markov Bloch QME formulation nor the NEGF formulation is capable 
of producing the LEPs. This inability of using LEPs for description of non-Markov evolution is quite general.
The concept of the Liouvillian exceptional points can be introduced only for Markovian dynamics.

We note that while the RWA should be used in derivation of the Bloch QME, within the NEGF
treatment the approximation may be relaxed. Such more general consideration will not affect the conclusions. 
Indeed, the inability to introduce LEPs directly follows from the fact that
the time-dependent characteristic expected for a LEP 
does not survive non-Markov time evolution of the system.
Absence of the RWA will only change a form of a time-dependent function convolution with which 
will destroy the expected LEP time dependence.
Similarly, as long as system evolution is non-Markov the conclusions hold for any driving frequency 
or in absence of external driving.

Finally, we stress that our work does not challenge existing experimental observations,
some of which are mentioned in introduction. We discuss theoretical treatments used for explanation 
of those experiments, and indicate possible pitfalls of the theory. 
For example, in many cases, theoretical treatments utilizing Markov 
description and employing exceptional points analysis will predict an abrupt `phase transition' when crossing
the exceptional point. In reality (i.e. within a more accurate theoretical analysis), the transition between 
two different regimes will be smooth. The importance of the difference between the two (approximate and more
accurate) theoretical descriptions and whether the approximate (Markov) treatment may lead to qualitative failures 
depends on the observable of interest.

\begin{acknowledgments}
This material is based upon work supported by the National Science Foundation under Grant No. 2154323.
\end{acknowledgments}
\appendix
\section{Derivation of Eq.(\ref{Sigma})}\label{appA}
We start from the definition of the single-particle Green's function, Eq.(\ref{Gij_def}).
Taking derivative in the first contour variable yields
\begin{equation}
\begin{split}
 & i\frac{\partial}{\partial\tau_1} G_{ij}(\tau_1,\tau_2) = \delta_{i,j}\,\delta(\tau_1,\tau_2)
 \\ &
 +\sum_{n=1,2}H^S_{in}(t_1)\, G_{nj}(\tau_1,\tau_2)
 \\ & -i\sum_{n,\alpha}\bigg[ 
 V_{ni,\alpha}\,\langle T_c\, \hat d_n(\tau_1)\hat b_\alpha(\tau_1)\, \hat d_j^\dagger(\tau_2)\rangle 
 \\ &\qquad\,\,\,
 +\langle T_c\, \hat b_\alpha^\dagger(\tau_1)\hat d_n(\tau_1)\,\hat d_j^\dagger(\tau_2)\rangle\, V_{\alpha,in} 
 \bigg]
 \end{split}
\end{equation} 
First order expansion of the scattering operator in the rightmost term of the expression leads to
\begin{equation}
\begin{split}
 & i\frac{\partial}{\partial\tau_1} G_{ij}(\tau_1,\tau_2) = \delta_{i,j}\,\delta(\tau_1,\tau_2)
 \\ &
 +\sum_{n=1,2}H^S_{in}(t_1)\, G_{nj}(\tau_1,\tau_2)
 -i\sum_{n,n_1,n_2}\int_c d\tau
 \\ &\quad
 \bigg[\,\, 
 \Pi_{ni,n_1n_2}(\tau_1,\tau)\, \langle T_c\, \hat d_n(\tau_1)\,\hat d^\dagger_{n_1}(\tau)\hat d_{n_2}(\tau)\, \hat d_j^\dagger(\tau_2)\rangle_0 
 \\ &\quad
 +\Pi_{n_1n_2,in}(\tau,\tau_1)\,
 \langle T_c\,\hat d_n(\tau_1)\, \hat d^\dagger_{n_2}(\tau)\hat d_{n_1}(\tau)\,\hat d_j^\dagger(\tau_2)\rangle_0
 \bigg]
 \end{split}
\end{equation} 
Here, $\Pi$ is defined in (\ref{Pi_def}) and subscript $0$ indicates evolution driven by system Hamiltonian.
Employing Wick's theorem to decouple multi-time correlation functions in the last term on the right side
and dressing the result yields the Hartree-Fock approximation, Eq.(\ref{Sigma}).

\section{Derivation of Eq.(\ref{rho_EOM})}\label{appB}
Here we derive exact EOM for density matrix, Eq.~(\ref{rho_EOM}),
starting from EOM for the Green's function (\ref{Gij_def}). 

We start with writing the left and right EOMs for the lesser projection of the 
Green's function (\ref{Gij_def})
\begin{widetext}
\begin{align}
\label{LEOM}
i\frac{\partial}{\partial t_1} G^{<}_{ij}(t_1,t_2) &= 
\varepsilon_i G^{<}_{ij}(t_1,t_2) - \mu E(t_1) G_{\bar i j}^{<}(t_1,t_2)
\\ &
+i\sum_{n,\alpha}\left[
V_{ni,\alpha}\langle \hat d_j^\dagger(t_2)\hat d_n(t_1)\hat b_\alpha(t_1)\rangle
+ V_{\alpha,in}\langle \hat d_j^\dagger(t_2)\hat d_n(t_1)\hat b^\dagger_\alpha(t_1)\rangle 
\right]
\nonumber 
\\
\label{REOM}
-i\frac{\partial}{\partial t_2} G^{<}_{ij}(t_1,t_2) &=
\varepsilon_j G^{<}_{ij}(t_1,t_2) - \mu E(t_2) G^{<}_{i\bar j}(t_1,t_2)
\\ &
+i\sum_{n,\alpha}\left[
V_{jn,\alpha}\langle \hat b_\alpha(t_2)\hat d_n^\dagger(t_2)\hat d_i(t_1)\rangle
+ V_{\alpha,nj}\langle \hat b_\alpha^\dagger(t_2)\hat d_n^\dagger(t_2)\hat d_i(t_1)\rangle
\right]
\nonumber
\end{align}
\end{widetext}
Taking $t_1=t_2\equiv t$ and subtracting (\ref{LEOM}) from (\ref{REOM}) yields
\begin{align}
\label{Glt_tt_EOM}
&-i\frac{d}{dt} G_{ij}^{<}(t,t) = \omega_{ji}G^{<}_{ij}(t,t) -
\mu E(t)\left[G^{<}_{i\bar j}(t,t)-G^{<}_{\bar i j}(t,t)\right]
\nonumber \\ &\qquad
+\sum_{n,\alpha}\left[
V_{ni,\alpha}\,{G}^{<}_{\alpha,nj}(t,t) 
+ {G}^{>}_{jn,\alpha}(t,t)\, V_{\alpha,in}
\right. \\ & \left.\qquad\qquad\,\,
- V_{jn,\alpha}\,{G}^{>}_{\alpha,in}(t,t) 
- {G}_{ni,\alpha}^{<}(t,t)\, V_{\alpha,nj}
\right]
\nonumber
\end{align}
Here,
\begin{equation}
\label{mixG_def}
\begin{split}
{G}_{\alpha,m_1m_2}(\tau_1,\tau_2) &\equiv
-i\bigg\langle T_c\, \hat b_\alpha(\tau_1)\left[\hat d_{m_1}^\dagger\hat d_{m_2}\right]^\dagger(\tau_2)\bigg\rangle
\\
{G}_{m_1m_2,\alpha}(\tau_1,\tau_2) &\equiv
-i\bigg\langle T_c\, \left[\hat d_{m_1}^\dagger\hat d_{m_2}\right](\tau_1)\,\,
\hat b_\alpha^\dagger(\tau_2)\bigg\rangle
\end{split}
\end{equation}
are the mixed system-bath Green's function which satisfy the Dyson equations
\begin{equation}
\label{calG_EOM}
\begin{split}
&{G}_{\alpha,m_1m_2}(\tau_1,\tau_2) =
\\ &\,\,\,\,\,
\sum_{n_1,n_2}\int_c d\tau'\, F^{(0)}_\alpha(\tau_1,\tau')\, V_{\alpha,n_1n_2}\,
G^{(2)}_{n_1n_2,m_1m_2}(\tau',\tau_2)
\\
&{G}_{m_1m_2,\alpha}(\tau_1,\tau_2) =
\\ &\,\,\,\,\,
\sum_{n_1,n_2}\int_c d\tau'\, G^{(2)}_{m_1m_2,n_1n_2}(\tau_1,\tau')\,
V_{n_1n_2,\alpha}\, F^{(0)}_\alpha(\tau',\tau_2) 
\end{split}
\end{equation}
Green's functions $F^{(0)}_\alpha$ and $G^{(2)}$ are defined in Eqs.~(\ref{F_def}) and (\ref{G2_def}), respectively.

Taking lesser and greater projections of (\ref{calG_EOM}),
setting $t_1=t_2\equiv t$, and substituting resulting expressions into 
(\ref{Glt_tt_EOM}) leads to (\ref{rho_EOM}).

\section{Derivation of Eq.(\ref{rho_Bloch})}\label{appC}
Here we derive Bloch equations, Eq.~(\ref{rho_Bloch}),
starting from exact EOM for density matrix, Eq.~(\ref{rho_EOM}).

Substituting the Kadanoff-Baym ansatz Eq.~(\ref{KB}) into (\ref{rho_EOM}) gives
\begin{widetext}
\begin{align}
\label{rho_dddd}
&\frac{d}{dt}\rho_{ij}(t) \approx i\omega_{ji}\rho_{ij}(t) 
- i\mu E(t)\left[\rho_{i\bar j}(t)-\rho_{\bar i j}(t)\right]
\\ & 
-i\sum_{n,n_1,n_2}\int_{-\infty}^{t}dt'\,\bigg[
\Pi_{ni,n_1n_2}^{>}(t-t')\,  e^{i\omega_{jn}(t-t')}\,
\left\langle\left[\hat d_{j}^\dagger\hat d_{n}\hat d_{n_1}^\dagger\hat d_{n_2}\right](t')\right\rangle
-\Pi_{ni,n_1n_2}^{<}(t-t')\, e^{i\omega_{jn}(t-t')}\,
\left\langle\left[\hat d_{n_1}^\dagger\hat d_{n_2}\hat d_j^\dagger\hat d_n\right](t')\right\rangle
\nonumber\\ &\qquad\qquad\qquad\quad\,\,\,
-\Pi_{jn,n_1n_2}^{>}(t-t')\,  e^{i\omega_{ni}(t-t')}
\left\langle\left[\hat d_{n}^\dagger\hat d_{i}\hat d_{n_1}^\dagger\hat d_{n_2}\right](t')\right\rangle
+\Pi_{jn,n_1n_2}^{<}(t-t')\, e^{i\omega_{ni}(t-t')}
\left\langle\left[\hat d_{n_1}^\dagger\hat d_{n_2}\hat d_{n}^\dagger\hat d_{i}\right](t')\right\rangle 
 \nonumber\\ &\qquad\qquad\qquad\quad\,\,\,
+\Pi_{n_1n_2,nj}^{>}(t'-t)\, e^{i\omega_{in}(t'-t)}\, 
\left\langle\left[\hat d_{n_2}^\dagger\hat d_{n_1}\hat d_{n}^\dagger\hat d_{i}\right](t')\right\rangle
-\Pi_{n_1n_2,nj}^{<}(t'-t)\, e^{i\omega_{in}(t'-t)}\, 
\left\langle\left[\hat d_{n}^\dagger\hat d_{i}\hat d_{n_2}^\dagger\hat d_{n_1}\right](t')\right\rangle
\nonumber\\ & \qquad\qquad\qquad\quad\,\,\,
-\Pi_{n_1n_2,in}^{>}(t'-t)\, e^{i\omega_{nj}(t'-t)}\, 
\left\langle\left[\hat d_{n_2}^\dagger\hat d_{n_1}\hat d_{j}^\dagger\hat d_{n}\right](t')\right\rangle
+\Pi_{n_1n_2,in}^{<}(t'-t)\, e^{i\omega_{nj}(t'-t)}\, 
\left\langle\left[\hat d_{j}^\dagger\hat d_{n}\hat d_{n_2}^\dagger\hat d_{n_1}\right](t')\right\rangle
\bigg]
\nonumber
\end{align}
\end{widetext}
where self-energy $\Pi$ is defined in Eq.~(\ref{Pi_def}).

In the single electron subspace of the problem 
\begin{equation}
\label{dddd}
\begin{split}
& \langle\hat d_1^\dagger\hat d_1\hat d_1^\dagger\hat d_1\rangle =
\langle\hat d_1^\dagger\hat d_2\hat d_2^\dagger\hat d_1\rangle =
\langle \hat d_1^\dagger\hat d_1\rangle \equiv \rho_{11},
\\
&\langle\hat d_2^\dagger\hat d_1\hat d_1^\dagger\hat d_1\rangle =
\langle\hat d_2^\dagger\hat d_2\hat d_2^\dagger\hat d_1\rangle =
\langle \hat d_2^\dagger\hat d_1\rangle \equiv \rho_{12}
\\
& \langle\hat d_1^\dagger\hat d_1\hat d_1^\dagger\hat d_2\rangle =
\langle\hat d_1^\dagger\hat d_2\hat d_2^\dagger\hat d_2\rangle =
\langle \hat d_1^\dagger\hat d_2\rangle \equiv \rho_{21}
\\
& \langle\hat d_2^\dagger\hat d_2\hat d_2^\dagger\hat d_2\rangle =
\langle\hat d_2^\dagger\hat d_1\hat d_1^\dagger\hat d_2\rangle =
\langle \hat d_2^\dagger\hat d_2\rangle \equiv \rho_{22}
\end{split}
\end{equation}
with all other averages zero.

Substituting (\ref{dddd}) in (\ref{rho_dddd}),
employing fast bath approximation
\begin{equation}
\begin{split}
& \rho_{11}(t')\approx\rho_{11}(t),\quad \rho_{12}(t')\approx \rho_{12}(t)\, e^{+i\omega_{21}(t'-t)},
\\
& \rho_{22}(t')\approx\rho_{22}(t),\quad \rho_{21}(t')\approx \rho_{21}(t)\, e^{-i\omega_{21}(t'-t)},
\end{split}
\end{equation}
and neglecting bath-induced couplings between populations and coherences
leads to (\ref{rho_Bloch}). \\*[0.7cm]

\section{Derivation of Eq.(\ref{GKB})}\label{appD}
Within the single-electron subspace of the problem there is a simple one-to-one correspondence between 
the single-particle and many-body states of the system. This correspondence
allows one to express the lesser and greater projections of the two-particle GF (\ref{G2_def}) as
\begin{equation}
\label{G2_Liouville}
\begin{split}
 &G^{(2)\, <}_{n_1n_2,n_3n_4}(t_1,t_2) \equiv -i\langle \left[\hat d_{n_3}^\dagger \hat d_{n_4}\right]^\dagger(t_2)\,\left[\hat d_{n_1}^\dagger\hat d_{n_2}\right](t_1)\rangle
 \\ &  \quad =
 -i\theta(t_1-t_2)\llangle n_2n_1\rvert\, \mathcal{U}(t_1,t_2)\,\lvert\,\rho_{SB}(t_2)\, n_4n_3 \rrangle
 \\ &\quad\,\,\,\,
 -i\theta(t_2-t_1)\llangle n_3n_4\rvert\, \mathcal{U}(t_2,t_1)\,\lvert\, n_1n_2\,\rho_{SB}(t_1) \rrangle
 \\
 &G^{(2)\, >}_{n_1n_2,n_3n_4}(t_1,t_2) \equiv -i\langle\left[\hat d_{n_1}^\dagger\hat d_{n_2}\right](t_1)\, \left[\hat d_{n_3}^\dagger \hat d_{n_4}\right]^\dagger(t_2)\rangle
 \\ & \quad =
 -i\theta(t_1-t_2)\llangle n_2n_1\rvert\, \mathcal{U}(t_1,t_2)\,\lvert\, n_4n_3\,\rho_{SB}(t_2) \rrangle
 \\ &\quad\,\,\,\,
 -i\theta(t_2-t_1)\llangle n_3n_4\rvert\, \mathcal{U}(t_2,t_1)\,\lvert\, \rho_{SB}(t_1)\, n_1n_2 \rrangle
\end{split}
\end{equation}
where rightmost sides of the expressions are written in the Liouville space notation
$\lvert n_1n_2\rrangle\equiv \lvert n_1\rangle\langle n_2\rvert$, 
$\hat \rho_{SB}$ is the total (system and bath) density operator, and
\begin{equation}
 \mathcal{U}(t_1,t_2)\equiv T\,\exp\left[-i\int_{t_2}^{t_1}dt\,\mathcal{L}(t)\right]
\end{equation}
is the Liouville space evolution operator.

By decoupling the system and the bath dynamics with projection super-operator
\begin{equation}
\label{proj}
\mathcal{P} \equiv \sum_{e,f} \lvert\, ef\,\rho_B^{eq}\,\rrangle\,\llangle\, ef\, I_B\rvert
\end{equation}
and introducing retarded and advanced Green's functions in Liouville space
\begin{equation}
\label{Gra_def}
\begin{split}
& \mathcal{G}^r_{n_1n_2,n_3n_4}(t_1,t_2) \equiv 
\\ &\quad
-i\theta(t_1-t_2)\llangle n_2n_1\, I_B\rvert\,  \mathcal{U}(t_1,t_2)\,
 \lvert n_4n_3\,\rho_B^{eq}\rrangle
 \\
&  \mathcal{G}^a_{n_1n_2,n_3n_4}(t_1,t_2) \equiv 
\\ &\quad
+i\theta(t_2-t_1)\llangle n_3n_4\, I_B\rvert\,  \mathcal{U}(t_1,t_2)\,
 \lvert n_1n_2\,\rho_B^{eq}\rrangle
\end{split}
\end{equation}
one can rewrite the exact expressions (\ref{G2_Liouville}) in the form of generalized Kadanoff-Baym ansatz in the 
Liouville space, Eq.(\ref{GKB}). 

Note that expressions for the Green's functions in the Liouville space given by
Eq.(\ref{Gra_Liouville}), are equivalent to the their definition (\ref{Gra_def}) via
\begin{equation}
\mathcal{U}_{eff}(t_1,t_2) \equiv \llangle I_B\rvert\, \mathcal{U}(t_1,t_2)\,\lvert\rho_B^{eq}\rrangle
\end{equation}


\begin{thebibliography}{60}%
\makeatletter
\providecommand \@ifxundefined [1]{%
 \@ifx{#1\undefined}
}%
\providecommand \@ifnum [1]{%
 \ifnum #1\expandafter \@firstoftwo
 \else \expandafter \@secondoftwo
 \fi
}%
\providecommand \@ifx [1]{%
 \ifx #1\expandafter \@firstoftwo
 \else \expandafter \@secondoftwo
 \fi
}%
\providecommand \natexlab [1]{#1}%
\providecommand \enquote  [1]{``#1''}%
\providecommand \bibnamefont  [1]{#1}%
\providecommand \bibfnamefont [1]{#1}%
\providecommand \citenamefont [1]{#1}%
\providecommand \href@noop [0]{\@secondoftwo}%
\providecommand \href [0]{\begingroup \@sanitize@url \@href}%
\providecommand \@href[1]{\@@startlink{#1}\@@href}%
\providecommand \@@href[1]{\endgroup#1\@@endlink}%
\providecommand \@sanitize@url [0]{\catcode `\\12\catcode `\$12\catcode
  `\&12\catcode `\#12\catcode `\^12\catcode `\_12\catcode `\%12\relax}%
\providecommand \@@startlink[1]{}%
\providecommand \@@endlink[0]{}%
\providecommand \url  [0]{\begingroup\@sanitize@url \@url }%
\providecommand \@url [1]{\endgroup\@href {#1}{\urlprefix }}%
\providecommand \urlprefix  [0]{URL }%
\providecommand \Eprint [0]{\href }%
\providecommand \doibase [0]{http://dx.doi.org/}%
\providecommand \selectlanguage [0]{\@gobble}%
\providecommand \bibinfo  [0]{\@secondoftwo}%
\providecommand \bibfield  [0]{\@secondoftwo}%
\providecommand \translation [1]{[#1]}%
\providecommand \BibitemOpen [0]{}%
\providecommand \bibitemStop [0]{}%
\providecommand \bibitemNoStop [0]{.\EOS\space}%
\providecommand \EOS [0]{\spacefactor3000\relax}%
\providecommand \BibitemShut  [1]{\csname bibitem#1\endcsname}%
\let\auto@bib@innerbib\@empty
\bibitem [{\citenamefont {Moiseyev}(2011)}]{moiseyev_non-hermitian_2011}%
  \BibitemOpen
  \bibfield  {author} {\bibinfo {author} {\bibfnamefont {N.}~\bibnamefont
  {Moiseyev}},\ }\href@noop {} {\emph {\bibinfo {title} {Non-{Hermitian}
  {Quantum} {Mechanics}}}}\ (\bibinfo  {publisher} {Cambridge University
  Press},\ \bibinfo {address} {Cambridge},\ \bibinfo {year} {2011})\BibitemShut
  {NoStop}%
\bibitem [{\citenamefont {Miri}\ and\ \citenamefont
  {Alù}(2019)}]{miri_exceptional_2019}%
  \BibitemOpen
  \bibfield  {author} {\bibinfo {author} {\bibfnamefont {M.-A.}\ \bibnamefont
  {Miri}}\ and\ \bibinfo {author} {\bibfnamefont {A.}~\bibnamefont {Alù}},\
  }\href {\doibase 10.1126/science.aar7709} {\bibfield  {journal} {\bibinfo
  {journal} {Science}\ }\textbf {\bibinfo {volume} {363}},\ \bibinfo {pages}
  {eaar7709} (\bibinfo {year} {2019})}\BibitemShut {NoStop}%
\bibitem [{\citenamefont {Lee}\ \emph {et~al.}(2009)\citenamefont {Lee},
  \citenamefont {Yang}, \citenamefont {Moon}, \citenamefont {Lee},
  \citenamefont {Shim}, \citenamefont {Kim}, \citenamefont {Lee},\ and\
  \citenamefont {An}}]{lee_observation_2009}%
  \BibitemOpen
  \bibfield  {author} {\bibinfo {author} {\bibfnamefont {S.-B.}\ \bibnamefont
  {Lee}}, \bibinfo {author} {\bibfnamefont {J.}~\bibnamefont {Yang}}, \bibinfo
  {author} {\bibfnamefont {S.}~\bibnamefont {Moon}}, \bibinfo {author}
  {\bibfnamefont {S.-Y.}\ \bibnamefont {Lee}}, \bibinfo {author} {\bibfnamefont
  {J.-B.}\ \bibnamefont {Shim}}, \bibinfo {author} {\bibfnamefont {S.~W.}\
  \bibnamefont {Kim}}, \bibinfo {author} {\bibfnamefont {J.-H.}\ \bibnamefont
  {Lee}}, \ and\ \bibinfo {author} {\bibfnamefont {K.}~\bibnamefont {An}},\
  }\href {\doibase 10.1103/PhysRevLett.103.134101} {\bibfield  {journal}
  {\bibinfo  {journal} {Phys. Rev. Lett.}\ }\textbf {\bibinfo {volume} {103}},\
  \bibinfo {pages} {134101} (\bibinfo {year} {2009})}\BibitemShut {NoStop}%
\bibitem [{\citenamefont {R{\" u}ter}\ \emph {et~al.}(2010)\citenamefont {R{\"
  u}ter}, \citenamefont {Makris}, \citenamefont {El-Ganainy}, \citenamefont
  {Christodoulides}, \citenamefont {Segev},\ and\ \citenamefont
  {Kip}}]{ruter_observation_2010}%
  \BibitemOpen
  \bibfield  {author} {\bibinfo {author} {\bibfnamefont {C.~E.}\ \bibnamefont
  {R{\" u}ter}}, \bibinfo {author} {\bibfnamefont {K.~G.}\ \bibnamefont
  {Makris}}, \bibinfo {author} {\bibfnamefont {R.}~\bibnamefont {El-Ganainy}},
  \bibinfo {author} {\bibfnamefont {D.~N.}\ \bibnamefont {Christodoulides}},
  \bibinfo {author} {\bibfnamefont {M.}~\bibnamefont {Segev}}, \ and\ \bibinfo
  {author} {\bibfnamefont {D.}~\bibnamefont {Kip}},\ }\href {\doibase
  10.1038/nphys1515} {\bibfield  {journal} {\bibinfo  {journal} {Nature Phys.}\
  }\textbf {\bibinfo {volume} {6}},\ \bibinfo {pages} {192} (\bibinfo {year}
  {2010})}\BibitemShut {NoStop}%
\bibitem [{\citenamefont {Hahn}\ \emph {et~al.}(2016)\citenamefont {Hahn},
  \citenamefont {Choi}, \citenamefont {Woong~Yoon}, \citenamefont {Ho~Song},
  \citenamefont {Hwan~Oh},\ and\ \citenamefont
  {Berini}}]{hahn_observation_2016}%
  \BibitemOpen
  \bibfield  {author} {\bibinfo {author} {\bibfnamefont {C.}~\bibnamefont
  {Hahn}}, \bibinfo {author} {\bibfnamefont {Y.}~\bibnamefont {Choi}}, \bibinfo
  {author} {\bibfnamefont {J.}~\bibnamefont {Woong~Yoon}}, \bibinfo {author}
  {\bibfnamefont {S.}~\bibnamefont {Ho~Song}}, \bibinfo {author} {\bibfnamefont
  {C.}~\bibnamefont {Hwan~Oh}}, \ and\ \bibinfo {author} {\bibfnamefont
  {P.}~\bibnamefont {Berini}},\ }\href {\doibase 10.1038/ncomms12201}
  {\bibfield  {journal} {\bibinfo  {journal} {Nat. Commun.}\ }\textbf {\bibinfo
  {volume} {7}},\ \bibinfo {pages} {12201} (\bibinfo {year}
  {2016})}\BibitemShut {NoStop}%
\bibitem [{\citenamefont {Doppler}\ \emph {et~al.}(2016)\citenamefont
  {Doppler}, \citenamefont {Mailybaev}, \citenamefont {B{\" o}hm},
  \citenamefont {Kuhl}, \citenamefont {Girschik}, \citenamefont {Libisch},
  \citenamefont {Milburn}, \citenamefont {Rabl}, \citenamefont {Moiseyev},\
  and\ \citenamefont {Rotter}}]{doppler_dynamically_2016}%
  \BibitemOpen
  \bibfield  {author} {\bibinfo {author} {\bibfnamefont {J.}~\bibnamefont
  {Doppler}}, \bibinfo {author} {\bibfnamefont {A.~A.}\ \bibnamefont
  {Mailybaev}}, \bibinfo {author} {\bibfnamefont {J.}~\bibnamefont {B{\"
  o}hm}}, \bibinfo {author} {\bibfnamefont {U.}~\bibnamefont {Kuhl}}, \bibinfo
  {author} {\bibfnamefont {A.}~\bibnamefont {Girschik}}, \bibinfo {author}
  {\bibfnamefont {F.}~\bibnamefont {Libisch}}, \bibinfo {author} {\bibfnamefont
  {T.~J.}\ \bibnamefont {Milburn}}, \bibinfo {author} {\bibfnamefont
  {P.}~\bibnamefont {Rabl}}, \bibinfo {author} {\bibfnamefont {N.}~\bibnamefont
  {Moiseyev}}, \ and\ \bibinfo {author} {\bibfnamefont {S.}~\bibnamefont
  {Rotter}},\ }\href {\doibase 10.1038/nature18605} {\bibfield  {journal}
  {\bibinfo  {journal} {Nature}\ }\textbf {\bibinfo {volume} {537}},\ \bibinfo
  {pages} {76} (\bibinfo {year} {2016})}\BibitemShut {NoStop}%
\bibitem [{\citenamefont {Ergoktas}\ \emph {et~al.}(2022)\citenamefont
  {Ergoktas}, \citenamefont {Soleymani}, \citenamefont {Kakenov}, \citenamefont
  {Wang}, \citenamefont {Smith}, \citenamefont {Bakan}, \citenamefont {Balci},
  \citenamefont {Principi}, \citenamefont {Novoselov}, \citenamefont
  {Ozdemir},\ and\ \citenamefont {Kocabas}}]{ergoktas_topological_2022}%
  \BibitemOpen
  \bibfield  {author} {\bibinfo {author} {\bibfnamefont {M.~S.}\ \bibnamefont
  {Ergoktas}}, \bibinfo {author} {\bibfnamefont {S.}~\bibnamefont {Soleymani}},
  \bibinfo {author} {\bibfnamefont {N.}~\bibnamefont {Kakenov}}, \bibinfo
  {author} {\bibfnamefont {K.}~\bibnamefont {Wang}}, \bibinfo {author}
  {\bibfnamefont {T.~B.}\ \bibnamefont {Smith}}, \bibinfo {author}
  {\bibfnamefont {G.}~\bibnamefont {Bakan}}, \bibinfo {author} {\bibfnamefont
  {S.}~\bibnamefont {Balci}}, \bibinfo {author} {\bibfnamefont
  {A.}~\bibnamefont {Principi}}, \bibinfo {author} {\bibfnamefont {K.~S.}\
  \bibnamefont {Novoselov}}, \bibinfo {author} {\bibfnamefont {S.~K.}\
  \bibnamefont {Ozdemir}}, \ and\ \bibinfo {author} {\bibfnamefont
  {C.}~\bibnamefont {Kocabas}},\ }\href {\doibase 10.1126/science.abn6528}
  {\bibfield  {journal} {\bibinfo  {journal} {Science}\ }\textbf {\bibinfo
  {volume} {376}},\ \bibinfo {pages} {184} (\bibinfo {year}
  {2022})}\BibitemShut {NoStop}%
\bibitem [{\citenamefont {Wu}\ \emph {et~al.}(2019)\citenamefont {Wu},
  \citenamefont {Liu}, \citenamefont {Geng}, \citenamefont {Song},
  \citenamefont {Ye}, \citenamefont {Duan}, \citenamefont {Rong},\ and\
  \citenamefont {Du}}]{wu_observation_2019}%
  \BibitemOpen
  \bibfield  {author} {\bibinfo {author} {\bibfnamefont {Y.}~\bibnamefont
  {Wu}}, \bibinfo {author} {\bibfnamefont {W.}~\bibnamefont {Liu}}, \bibinfo
  {author} {\bibfnamefont {J.}~\bibnamefont {Geng}}, \bibinfo {author}
  {\bibfnamefont {X.}~\bibnamefont {Song}}, \bibinfo {author} {\bibfnamefont
  {X.}~\bibnamefont {Ye}}, \bibinfo {author} {\bibfnamefont {C.-K.}\
  \bibnamefont {Duan}}, \bibinfo {author} {\bibfnamefont {X.}~\bibnamefont
  {Rong}}, \ and\ \bibinfo {author} {\bibfnamefont {J.}~\bibnamefont {Du}},\
  }\href {\doibase 10.1126/science.aaw8205} {\bibfield  {journal} {\bibinfo
  {journal} {Science}\ }\textbf {\bibinfo {volume} {364}},\ \bibinfo {pages}
  {878} (\bibinfo {year} {2019})}\BibitemShut {NoStop}%
\bibitem [{\citenamefont {Wiersig}(2020)}]{wiersig_review_2020}%
  \BibitemOpen
  \bibfield  {author} {\bibinfo {author} {\bibfnamefont {J.}~\bibnamefont
  {Wiersig}},\ }\href {\doibase 10.1364/PRJ.396115} {\bibfield  {journal}
  {\bibinfo  {journal} {Photon. Res.}\ }\textbf {\bibinfo {volume} {8}},\
  \bibinfo {pages} {1457} (\bibinfo {year} {2020})}\BibitemShut {NoStop}%
\bibitem [{\citenamefont {Yang}\ \emph {et~al.}(2023)\citenamefont {Yang},
  \citenamefont {Zhang}, \citenamefont {Liao}, \citenamefont {Liu},
  \citenamefont {Zhou}, \citenamefont {Zhou}, \citenamefont {Xu}, \citenamefont
  {Han}, \citenamefont {Li},\ and\ \citenamefont {Guo}}]{mu_realization_2023}%
  \BibitemOpen
  \bibfield  {author} {\bibinfo {author} {\bibfnamefont {M.}~\bibnamefont
  {Yang}}, \bibinfo {author} {\bibfnamefont {H.-Q.}\ \bibnamefont {Zhang}},
  \bibinfo {author} {\bibfnamefont {Y.-W.}\ \bibnamefont {Liao}}, \bibinfo
  {author} {\bibfnamefont {Z.-H.}\ \bibnamefont {Liu}}, \bibinfo {author}
  {\bibfnamefont {Z.-W.}\ \bibnamefont {Zhou}}, \bibinfo {author}
  {\bibfnamefont {X.-X.}\ \bibnamefont {Zhou}}, \bibinfo {author}
  {\bibfnamefont {J.-S.}\ \bibnamefont {Xu}}, \bibinfo {author} {\bibfnamefont
  {Y.-J.}\ \bibnamefont {Han}}, \bibinfo {author} {\bibfnamefont {C.-F.}\
  \bibnamefont {Li}}, \ and\ \bibinfo {author} {\bibfnamefont {G.-C.}\
  \bibnamefont {Guo}},\ }\href {\doibase 10.1126/sciadv.abp8943} {\bibfield
  {journal} {\bibinfo  {journal} {Science Advances}\ }\textbf {\bibinfo
  {volume} {9}},\ \bibinfo {pages} {eabp8943} (\bibinfo {year}
  {2023})}\BibitemShut {NoStop}%
\bibitem [{\citenamefont {Liang}\ \emph {et~al.}(2023)\citenamefont {Liang},
  \citenamefont {Tang}, \citenamefont {Xu},\ and\ \citenamefont
  {Liu}}]{liang_observation_2023}%
  \BibitemOpen
  \bibfield  {author} {\bibinfo {author} {\bibfnamefont {C.}~\bibnamefont
  {Liang}}, \bibinfo {author} {\bibfnamefont {Y.}~\bibnamefont {Tang}},
  \bibinfo {author} {\bibfnamefont {A.-N.}\ \bibnamefont {Xu}}, \ and\ \bibinfo
  {author} {\bibfnamefont {Y.-C.}\ \bibnamefont {Liu}},\ }\href {\doibase
  10.1103/PhysRevLett.130.263601} {\bibfield  {journal} {\bibinfo  {journal}
  {Phys. Rev. Lett.}\ }\textbf {\bibinfo {volume} {130}},\ \bibinfo {pages}
  {263601} (\bibinfo {year} {2023})}\BibitemShut {NoStop}%
\bibitem [{\citenamefont {Zhang}\ \emph {et~al.}(2018)\citenamefont {Zhang},
  \citenamefont {Peng}, \citenamefont {{\" O}zdemir}, \citenamefont {Pichler},
  \citenamefont {Krimer}, \citenamefont {Zhao}, \citenamefont {Nori},
  \citenamefont {Liu}, \citenamefont {Rotter},\ and\ \citenamefont
  {Yang}}]{zhang_a_2018}%
  \BibitemOpen
  \bibfield  {author} {\bibinfo {author} {\bibfnamefont {J.}~\bibnamefont
  {Zhang}}, \bibinfo {author} {\bibfnamefont {B.}~\bibnamefont {Peng}},
  \bibinfo {author} {\bibfnamefont {{\c S}.~K.}\ \bibnamefont {{\" O}zdemir}},
  \bibinfo {author} {\bibfnamefont {K.}~\bibnamefont {Pichler}}, \bibinfo
  {author} {\bibfnamefont {D.~O.}\ \bibnamefont {Krimer}}, \bibinfo {author}
  {\bibfnamefont {G.}~\bibnamefont {Zhao}}, \bibinfo {author} {\bibfnamefont
  {F.}~\bibnamefont {Nori}}, \bibinfo {author} {\bibfnamefont {Y.-x.}\
  \bibnamefont {Liu}}, \bibinfo {author} {\bibfnamefont {S.}~\bibnamefont
  {Rotter}}, \ and\ \bibinfo {author} {\bibfnamefont {L.}~\bibnamefont
  {Yang}},\ }\href {\doibase 10.1038/s41566-018-0213-5} {\bibfield  {journal}
  {\bibinfo  {journal} {Nat. Photonics}\ }\textbf {\bibinfo {volume} {12}},\
  \bibinfo {pages} {479} (\bibinfo {year} {2018})}\BibitemShut {NoStop}%
\bibitem [{\citenamefont {Naghiloo}\ \emph {et~al.}(2019)\citenamefont
  {Naghiloo}, \citenamefont {Abbasi}, \citenamefont {Joglekar},\ and\
  \citenamefont {Murch}}]{naghiloo_quantum_2019}%
  \BibitemOpen
  \bibfield  {author} {\bibinfo {author} {\bibfnamefont {M.}~\bibnamefont
  {Naghiloo}}, \bibinfo {author} {\bibfnamefont {M.}~\bibnamefont {Abbasi}},
  \bibinfo {author} {\bibfnamefont {Y.~N.}\ \bibnamefont {Joglekar}}, \ and\
  \bibinfo {author} {\bibfnamefont {K.~W.}\ \bibnamefont {Murch}},\ }\href
  {\doibase 10.1038/s41567-019-0652-z} {\bibfield  {journal} {\bibinfo
  {journal} {Nature Phys.}\ }\textbf {\bibinfo {volume} {15}},\ \bibinfo
  {pages} {1232} (\bibinfo {year} {2019})}\BibitemShut {NoStop}%
\bibitem [{\citenamefont {Gao}\ \emph {et~al.}(2015)\citenamefont {Gao},
  \citenamefont {Estrecho}, \citenamefont {Bliokh}, \citenamefont {Liew},
  \citenamefont {Fraser}, \citenamefont {Brodbeck}, \citenamefont {Kamp},
  \citenamefont {Schneider}, \citenamefont {H{\" o}fling}, \citenamefont
  {Yamamoto}, \citenamefont {Nori}, \citenamefont {Kivshar}, \citenamefont
  {Truscott}, \citenamefont {G.},\ and\ \citenamefont
  {Ostrovskaya}}]{gao_observation_2015}%
  \BibitemOpen
  \bibfield  {author} {\bibinfo {author} {\bibfnamefont {T.}~\bibnamefont
  {Gao}}, \bibinfo {author} {\bibfnamefont {E.}~\bibnamefont {Estrecho}},
  \bibinfo {author} {\bibfnamefont {K.~Y.}\ \bibnamefont {Bliokh}}, \bibinfo
  {author} {\bibfnamefont {T.~C.~H.}\ \bibnamefont {Liew}}, \bibinfo {author}
  {\bibfnamefont {M.~D.}\ \bibnamefont {Fraser}}, \bibinfo {author}
  {\bibfnamefont {M.}~\bibnamefont {Brodbeck}}, \bibinfo {author}
  {\bibfnamefont {M.}~\bibnamefont {Kamp}}, \bibinfo {author} {\bibfnamefont
  {C.}~\bibnamefont {Schneider}}, \bibinfo {author} {\bibfnamefont
  {S.}~\bibnamefont {H{\" o}fling}}, \bibinfo {author} {\bibfnamefont
  {Y.}~\bibnamefont {Yamamoto}}, \bibinfo {author} {\bibfnamefont
  {F.}~\bibnamefont {Nori}}, \bibinfo {author} {\bibfnamefont {Y.~S.}\
  \bibnamefont {Kivshar}}, \bibinfo {author} {\bibfnamefont {A.~G.}\
  \bibnamefont {Truscott}}, \bibinfo {author} {\bibfnamefont {D.~R.}\
  \bibnamefont {G.}}, \ and\ \bibinfo {author} {\bibfnamefont {N.~A.}\
  \bibnamefont {Ostrovskaya}},\ }\href {\doibase 10.1038/nature15522}
  {\bibfield  {journal} {\bibinfo  {journal} {Nature}\ }\textbf {\bibinfo
  {volume} {526}},\ \bibinfo {pages} {554} (\bibinfo {year}
  {2015})}\BibitemShut {NoStop}%
\bibitem [{\citenamefont {Xu}\ \emph {et~al.}(2023)\citenamefont {Xu},
  \citenamefont {Zhou}, \citenamefont {Li}, \citenamefont {Cao}, \citenamefont
  {Chen}, \citenamefont {Xiao}, \citenamefont {Yang},\ and\ \citenamefont
  {Qiu}}]{xu_non-hermitian_2023}%
  \BibitemOpen
  \bibfield  {author} {\bibinfo {author} {\bibfnamefont {G.}~\bibnamefont
  {Xu}}, \bibinfo {author} {\bibfnamefont {X.}~\bibnamefont {Zhou}}, \bibinfo
  {author} {\bibfnamefont {Y.}~\bibnamefont {Li}}, \bibinfo {author}
  {\bibfnamefont {Q.}~\bibnamefont {Cao}}, \bibinfo {author} {\bibfnamefont
  {W.}~\bibnamefont {Chen}}, \bibinfo {author} {\bibfnamefont {Y.}~\bibnamefont
  {Xiao}}, \bibinfo {author} {\bibfnamefont {L.}~\bibnamefont {Yang}}, \ and\
  \bibinfo {author} {\bibfnamefont {C.-W.}\ \bibnamefont {Qiu}},\ }\href
  {\doibase 10.1103/PhysRevLett.130.266303} {\bibfield  {journal} {\bibinfo
  {journal} {Phys. Rev. Lett.}\ }\textbf {\bibinfo {volume} {130}},\ \bibinfo
  {pages} {266303} (\bibinfo {year} {2023})}\BibitemShut {NoStop}%
\bibitem [{\citenamefont {G{\" u}nther}\ \emph {et~al.}(2007)\citenamefont
  {G{\" u}nther}, \citenamefont {Rotter},\ and\ \citenamefont
  {Samsonov}}]{gunther_projective_2007}%
  \BibitemOpen
  \bibfield  {author} {\bibinfo {author} {\bibfnamefont {U.}~\bibnamefont {G{\"
  u}nther}}, \bibinfo {author} {\bibfnamefont {I.}~\bibnamefont {Rotter}}, \
  and\ \bibinfo {author} {\bibfnamefont {B.~F.}\ \bibnamefont {Samsonov}},\
  }\href {http://stacks.iop.org/1751-8121/40/i=30/a=014} {\bibfield  {journal}
  {\bibinfo  {journal} {J. Phys. A}\ }\textbf {\bibinfo {volume} {40}},\
  \bibinfo {pages} {8815} (\bibinfo {year} {2007})}\BibitemShut {NoStop}%
\bibitem [{\citenamefont {Rotter}(2009)}]{rotter_non-hermitian_2009}%
  \BibitemOpen
  \bibfield  {author} {\bibinfo {author} {\bibfnamefont {I.}~\bibnamefont
  {Rotter}},\ }\href {http://stacks.iop.org/1751-8121/42/i=15/a=153001}
  {\bibfield  {journal} {\bibinfo  {journal} {J. Phys. A}\ }\textbf {\bibinfo
  {volume} {42}},\ \bibinfo {pages} {153001} (\bibinfo {year}
  {2009})}\BibitemShut {NoStop}%
\bibitem [{\citenamefont {Toroker}\ and\ \citenamefont
  {Peskin}(2009)}]{toroker_relation_2009}%
  \BibitemOpen
  \bibfield  {author} {\bibinfo {author} {\bibfnamefont {M.~C.}\ \bibnamefont
  {Toroker}}\ and\ \bibinfo {author} {\bibfnamefont {U.}~\bibnamefont
  {Peskin}},\ }\href {http://stacks.iop.org/0953-4075/42/i=4/a=044013}
  {\bibfield  {journal} {\bibinfo  {journal} {J. Phys. B}\ }\textbf {\bibinfo
  {volume} {42}},\ \bibinfo {pages} {044013} (\bibinfo {year}
  {2009})}\BibitemShut {NoStop}%
\bibitem [{\citenamefont {Uzdin}\ \emph {et~al.}(2011)\citenamefont {Uzdin},
  \citenamefont {Mailybaev},\ and\ \citenamefont
  {Moiseyev}}]{uzdin_observability_2011}%
  \BibitemOpen
  \bibfield  {author} {\bibinfo {author} {\bibfnamefont {R.}~\bibnamefont
  {Uzdin}}, \bibinfo {author} {\bibfnamefont {A.}~\bibnamefont {Mailybaev}}, \
  and\ \bibinfo {author} {\bibfnamefont {N.}~\bibnamefont {Moiseyev}},\ }\href
  {\doibase 10.1088/1751-8113/44/43/435302} {\bibfield  {journal} {\bibinfo
  {journal} {J. Phys. A: Math. and Theor.}\ }\textbf {\bibinfo {volume} {44}},\
  \bibinfo {pages} {435302} (\bibinfo {year} {2011})}\BibitemShut {NoStop}%
\bibitem [{\citenamefont {Heiss}(2012)}]{heiss_physics_2012}%
  \BibitemOpen
  \bibfield  {author} {\bibinfo {author} {\bibfnamefont {W.~D.}\ \bibnamefont
  {Heiss}},\ }\href {\doibase 10.1088/1751-8113/45/44/444016} {\bibfield
  {journal} {\bibinfo  {journal} {J. Phys. A}\ }\textbf {\bibinfo {volume}
  {45}},\ \bibinfo {pages} {444016} (\bibinfo {year} {2012})}\BibitemShut
  {NoStop}%
\bibitem [{\citenamefont {Garmon}\ \emph {et~al.}(2012)\citenamefont {Garmon},
  \citenamefont {Rotter}, \citenamefont {Hatano},\ and\ \citenamefont
  {Segal}}]{garmon_analysis_2012}%
  \BibitemOpen
  \bibfield  {author} {\bibinfo {author} {\bibfnamefont {S.}~\bibnamefont
  {Garmon}}, \bibinfo {author} {\bibfnamefont {I.}~\bibnamefont {Rotter}},
  \bibinfo {author} {\bibfnamefont {N.}~\bibnamefont {Hatano}}, \ and\ \bibinfo
  {author} {\bibfnamefont {D.}~\bibnamefont {Segal}},\ }\href {\doibase
  10.1007/s10773-012-1240-5} {\bibfield  {journal} {\bibinfo  {journal} {Int.
  J. Theor. Phys.}\ }\textbf {\bibinfo {volume} {51}},\ \bibinfo {pages} {3536}
  (\bibinfo {year} {2012})}\BibitemShut {NoStop}%
\bibitem [{\citenamefont {Delga}\ \emph {et~al.}(2014)\citenamefont {Delga},
  \citenamefont {Feist}, \citenamefont {Bravo-Abad},\ and\ \citenamefont
  {Garcia-Vidal}}]{delga_theory_2014}%
  \BibitemOpen
  \bibfield  {author} {\bibinfo {author} {\bibfnamefont {A.}~\bibnamefont
  {Delga}}, \bibinfo {author} {\bibfnamefont {J.}~\bibnamefont {Feist}},
  \bibinfo {author} {\bibfnamefont {J.}~\bibnamefont {Bravo-Abad}}, \ and\
  \bibinfo {author} {\bibfnamefont {F.~J.}\ \bibnamefont {Garcia-Vidal}},\
  }\href {\doibase 10.1088/2040-8978/16/11/114018} {\bibfield  {journal}
  {\bibinfo  {journal} {Journal of Optics}\ }\textbf {\bibinfo {volume} {16}},\
  \bibinfo {pages} {114018} (\bibinfo {year} {2014})}\BibitemShut {NoStop}%
\bibitem [{\citenamefont {Rotter}\ and\ \citenamefont
  {Bird}(2015)}]{rotter_review_2015}%
  \BibitemOpen
  \bibfield  {author} {\bibinfo {author} {\bibfnamefont {I.}~\bibnamefont
  {Rotter}}\ and\ \bibinfo {author} {\bibfnamefont {J.~P.}\ \bibnamefont
  {Bird}},\ }\href {http://stacks.iop.org/0034-4885/78/i=11/a=114001}
  {\bibfield  {journal} {\bibinfo  {journal} {Rep. Prog. Phys.}\ }\textbf
  {\bibinfo {volume} {78}},\ \bibinfo {pages} {114001} (\bibinfo {year}
  {2015})}\BibitemShut {NoStop}%
\bibitem [{\citenamefont {Yang}\ \emph {et~al.}(2020)\citenamefont {Yang},
  \citenamefont {Wei}, \citenamefont {Sheng},\ and\ \citenamefont
  {Wu}}]{yang_phonon_2020}%
  \BibitemOpen
  \bibfield  {author} {\bibinfo {author} {\bibfnamefont {C.}~\bibnamefont
  {Yang}}, \bibinfo {author} {\bibfnamefont {X.}~\bibnamefont {Wei}}, \bibinfo
  {author} {\bibfnamefont {J.}~\bibnamefont {Sheng}}, \ and\ \bibinfo {author}
  {\bibfnamefont {H.}~\bibnamefont {Wu}},\ }\href {\doibase
  10.1038/s41467-020-18426-4} {\bibfield  {journal} {\bibinfo  {journal} {Nat.
  Commun.}\ }\textbf {\bibinfo {volume} {11}},\ \bibinfo {pages} {4656}
  (\bibinfo {year} {2020})}\BibitemShut {NoStop}%
\bibitem [{\citenamefont {Engelhardt}\ and\ \citenamefont
  {Cao}(2022)}]{engelhardt_unusual_2022}%
  \BibitemOpen
  \bibfield  {author} {\bibinfo {author} {\bibfnamefont {G.}~\bibnamefont
  {Engelhardt}}\ and\ \bibinfo {author} {\bibfnamefont {J.}~\bibnamefont
  {Cao}},\ }\href {\doibase 10.1103/PhysRevB.105.064205} {\bibfield  {journal}
  {\bibinfo  {journal} {Phys. Rev. B}\ }\textbf {\bibinfo {volume} {105}},\
  \bibinfo {pages} {064205} (\bibinfo {year} {2022})}\BibitemShut {NoStop}%
\bibitem [{\citenamefont {Ferrier}\ \emph {et~al.}(2022)\citenamefont
  {Ferrier}, \citenamefont {Bouteyre}, \citenamefont {Pick}, \citenamefont
  {Cueff}, \citenamefont {Dang}, \citenamefont {Diederichs}, \citenamefont
  {Belarouci}, \citenamefont {Benyattou}, \citenamefont {Zhao}, \citenamefont
  {Su}, \citenamefont {Xing}, \citenamefont {Xiong},\ and\ \citenamefont
  {Nguyen}}]{ferrier_unveiling_2022}%
  \BibitemOpen
  \bibfield  {author} {\bibinfo {author} {\bibfnamefont {L.}~\bibnamefont
  {Ferrier}}, \bibinfo {author} {\bibfnamefont {P.}~\bibnamefont {Bouteyre}},
  \bibinfo {author} {\bibfnamefont {A.}~\bibnamefont {Pick}}, \bibinfo {author}
  {\bibfnamefont {S.}~\bibnamefont {Cueff}}, \bibinfo {author} {\bibfnamefont
  {N.}~\bibnamefont {Dang}}, \bibinfo {author} {\bibfnamefont {C.}~\bibnamefont
  {Diederichs}}, \bibinfo {author} {\bibfnamefont {A.}~\bibnamefont
  {Belarouci}}, \bibinfo {author} {\bibfnamefont {T.}~\bibnamefont
  {Benyattou}}, \bibinfo {author} {\bibfnamefont {J.}~\bibnamefont {Zhao}},
  \bibinfo {author} {\bibfnamefont {R.}~\bibnamefont {Su}}, \bibinfo {author}
  {\bibfnamefont {J.}~\bibnamefont {Xing}}, \bibinfo {author} {\bibfnamefont
  {Q.}~\bibnamefont {Xiong}}, \ and\ \bibinfo {author} {\bibfnamefont
  {H.}~\bibnamefont {Nguyen}},\ }\href {\doibase
  10.1103/PhysRevLett.129.083602} {\bibfield  {journal} {\bibinfo  {journal}
  {Phys. Rev. Lett.}\ }\textbf {\bibinfo {volume} {129}},\ \bibinfo {pages}
  {083602} (\bibinfo {year} {2022})}\BibitemShut {NoStop}%
\bibitem [{\citenamefont {Engelhardt}\ and\ \citenamefont
  {Cao}(2023)}]{engelhardt_polariton_2023}%
  \BibitemOpen
  \bibfield  {author} {\bibinfo {author} {\bibfnamefont {G.}~\bibnamefont
  {Engelhardt}}\ and\ \bibinfo {author} {\bibfnamefont {J.}~\bibnamefont
  {Cao}},\ }\href {\doibase 10.1103/PhysRevLett.130.213602} {\bibfield
  {journal} {\bibinfo  {journal} {Phys. Rev. Lett.}\ }\textbf {\bibinfo
  {volume} {130}},\ \bibinfo {pages} {213602} (\bibinfo {year}
  {2023})}\BibitemShut {NoStop}%
\bibitem [{\citenamefont {Li}\ \emph {et~al.}(2023)\citenamefont {Li},
  \citenamefont {Chen}, \citenamefont {Abbasi}, \citenamefont {Murch},\ and\
  \citenamefont {Whaley}}]{li_speeding_2023}%
  \BibitemOpen
  \bibfield  {author} {\bibinfo {author} {\bibfnamefont {Z.-Z.}\ \bibnamefont
  {Li}}, \bibinfo {author} {\bibfnamefont {W.}~\bibnamefont {Chen}}, \bibinfo
  {author} {\bibfnamefont {M.}~\bibnamefont {Abbasi}}, \bibinfo {author}
  {\bibfnamefont {K.~W.}\ \bibnamefont {Murch}}, \ and\ \bibinfo {author}
  {\bibfnamefont {K.~B.}\ \bibnamefont {Whaley}},\ }\href {\doibase
  10.1103/PhysRevLett.131.100202} {\bibfield  {journal} {\bibinfo  {journal}
  {Phys. Rev. Lett.}\ }\textbf {\bibinfo {volume} {131}},\ \bibinfo {pages}
  {100202} (\bibinfo {year} {2023})}\BibitemShut {NoStop}%
\bibitem [{\citenamefont {Pick}\ \emph {et~al.}(2019)\citenamefont {Pick},
  \citenamefont {Silberstein}, \citenamefont {Moiseyev},\ and\ \citenamefont
  {Bar-Gill}}]{pick_robust_2019}%
  \BibitemOpen
  \bibfield  {author} {\bibinfo {author} {\bibfnamefont {A.}~\bibnamefont
  {Pick}}, \bibinfo {author} {\bibfnamefont {S.}~\bibnamefont {Silberstein}},
  \bibinfo {author} {\bibfnamefont {N.}~\bibnamefont {Moiseyev}}, \ and\
  \bibinfo {author} {\bibfnamefont {N.}~\bibnamefont {Bar-Gill}},\ }\href
  {\doibase 10.1103/PhysRevResearch.1.013015} {\bibfield  {journal} {\bibinfo
  {journal} {Phys. Rev. Res.}\ }\textbf {\bibinfo {volume} {1}},\ \bibinfo
  {pages} {013015} (\bibinfo {year} {2019})}\BibitemShut {NoStop}%
\bibitem [{\citenamefont {Minganti}\ \emph {et~al.}(2019)\citenamefont
  {Minganti}, \citenamefont {Miranowicz}, \citenamefont {Chhajlany},\ and\
  \citenamefont {Nori}}]{minganti_quantum_2019}%
  \BibitemOpen
  \bibfield  {author} {\bibinfo {author} {\bibfnamefont {F.}~\bibnamefont
  {Minganti}}, \bibinfo {author} {\bibfnamefont {A.}~\bibnamefont
  {Miranowicz}}, \bibinfo {author} {\bibfnamefont {R.~W.}\ \bibnamefont
  {Chhajlany}}, \ and\ \bibinfo {author} {\bibfnamefont {F.}~\bibnamefont
  {Nori}},\ }\href {\doibase 10.1103/PhysRevA.100.062131} {\bibfield  {journal}
  {\bibinfo  {journal} {Phys. Rev. A}\ }\textbf {\bibinfo {volume} {100}},\
  \bibinfo {pages} {062131} (\bibinfo {year} {2019})}\BibitemShut {NoStop}%
\bibitem [{\citenamefont {Minganti}\ \emph {et~al.}(2020)\citenamefont
  {Minganti}, \citenamefont {Miranowicz}, \citenamefont {Chhajlany},
  \citenamefont {Arkhipov},\ and\ \citenamefont
  {Nori}}]{minganti_hybrid-liouvillian_2020}%
  \BibitemOpen
  \bibfield  {author} {\bibinfo {author} {\bibfnamefont {F.}~\bibnamefont
  {Minganti}}, \bibinfo {author} {\bibfnamefont {A.}~\bibnamefont
  {Miranowicz}}, \bibinfo {author} {\bibfnamefont {R.~W.}\ \bibnamefont
  {Chhajlany}}, \bibinfo {author} {\bibfnamefont {I.~I.}\ \bibnamefont
  {Arkhipov}}, \ and\ \bibinfo {author} {\bibfnamefont {F.}~\bibnamefont
  {Nori}},\ }\href {\doibase 10.1103/PhysRevA.101.062112} {\bibfield  {journal}
  {\bibinfo  {journal} {Phys. Rev. A}\ }\textbf {\bibinfo {volume} {101}},\
  \bibinfo {pages} {062112} (\bibinfo {year} {2020})}\BibitemShut {NoStop}%
\bibitem [{\citenamefont {Arkhipov}\ \emph {et~al.}(2020)\citenamefont
  {Arkhipov}, \citenamefont {Miranowicz}, \citenamefont {Minganti},\ and\
  \citenamefont {Nori}}]{arkhipov_liouvillian_2020}%
  \BibitemOpen
  \bibfield  {author} {\bibinfo {author} {\bibfnamefont {I.~I.}\ \bibnamefont
  {Arkhipov}}, \bibinfo {author} {\bibfnamefont {A.}~\bibnamefont
  {Miranowicz}}, \bibinfo {author} {\bibfnamefont {F.}~\bibnamefont
  {Minganti}}, \ and\ \bibinfo {author} {\bibfnamefont {F.}~\bibnamefont
  {Nori}},\ }\href {\doibase 10.1103/PhysRevA.102.033715} {\bibfield  {journal}
  {\bibinfo  {journal} {Phys. Rev. A}\ }\textbf {\bibinfo {volume} {102}},\
  \bibinfo {pages} {033715} (\bibinfo {year} {2020})}\BibitemShut {NoStop}%
\bibitem [{\citenamefont {Chimzak}\ \emph {et~al.}(2023)\citenamefont
  {Chimzak}, \citenamefont {Kowalewska-Kudlaszyk}, \citenamefont {Lange},
  \citenamefont {Bartkiewicz},\ and\ \citenamefont {Pei{\v
  r}ina~Jr.}}]{chimczak_the_2023}%
  \BibitemOpen
  \bibfield  {author} {\bibinfo {author} {\bibfnamefont {G.}~\bibnamefont
  {Chimzak}}, \bibinfo {author} {\bibfnamefont {A.}~\bibnamefont
  {Kowalewska-Kudlaszyk}}, \bibinfo {author} {\bibfnamefont {E.}~\bibnamefont
  {Lange}}, \bibinfo {author} {\bibfnamefont {K.}~\bibnamefont {Bartkiewicz}},
  \ and\ \bibinfo {author} {\bibfnamefont {J.}~\bibnamefont {Pei{\v
  r}ina~Jr.}},\ }\href {\doibase 10.1038/s41598-023-32864-2} {\bibfield
  {journal} {\bibinfo  {journal} {Sci. Rep.}\ }\textbf {\bibinfo {volume}
  {13}},\ \bibinfo {pages} {5859} (\bibinfo {year} {2023})}\BibitemShut
  {NoStop}%
\bibitem [{\citenamefont {Hashimoto}\ \emph
  {et~al.}(2016{\natexlab{a}})\citenamefont {Hashimoto}, \citenamefont {Kanki},
  \citenamefont {Garmon}, \citenamefont {Tanaka},\ and\ \citenamefont
  {Petrosky}}]{hashimoto_on_2016}%
  \BibitemOpen
  \bibfield  {author} {\bibinfo {author} {\bibfnamefont {K.}~\bibnamefont
  {Hashimoto}}, \bibinfo {author} {\bibfnamefont {K.}~\bibnamefont {Kanki}},
  \bibinfo {author} {\bibfnamefont {S.}~\bibnamefont {Garmon}}, \bibinfo
  {author} {\bibfnamefont {S.}~\bibnamefont {Tanaka}}, \ and\ \bibinfo {author}
  {\bibfnamefont {T.}~\bibnamefont {Petrosky}},\ }\href {\doibase
  10.1093/ptep/ptw039} {\bibfield  {journal} {\bibinfo  {journal} {Progress of
  Theoretical and Experimental Physics}\ }\textbf {\bibinfo {volume} {2016}},\
  \bibinfo {pages} {053A02} (\bibinfo {year} {2016}{\natexlab{a}})}\BibitemShut
  {NoStop}%
\bibitem [{\citenamefont {Hashimoto}\ \emph
  {et~al.}(2016{\natexlab{b}})\citenamefont {Hashimoto}, \citenamefont {Kanki},
  \citenamefont {Tanaka},\ and\ \citenamefont
  {Petrosky}}]{hashimoto_physical_2016}%
  \BibitemOpen
  \bibfield  {author} {\bibinfo {author} {\bibfnamefont {K.}~\bibnamefont
  {Hashimoto}}, \bibinfo {author} {\bibfnamefont {K.}~\bibnamefont {Kanki}},
  \bibinfo {author} {\bibfnamefont {S.}~\bibnamefont {Tanaka}}, \ and\ \bibinfo
  {author} {\bibfnamefont {T.}~\bibnamefont {Petrosky}},\ }in\ \href@noop {}
  {\emph {\bibinfo {booktitle} {Non-Hermitian Hamiltonians in Quantum
  Physics}}},\ \bibinfo {editor} {edited by\ \bibinfo {editor} {\bibfnamefont
  {F.}~\bibnamefont {Bagarello}}, \bibinfo {editor} {\bibfnamefont
  {R.}~\bibnamefont {Passante}}, \ and\ \bibinfo {editor} {\bibfnamefont
  {C.}~\bibnamefont {Trapani}}}\ (\bibinfo  {publisher} {Springer International
  Publishing},\ \bibinfo {address} {Cham},\ \bibinfo {year} {2016})\ pp.\
  \bibinfo {pages} {263--279}\BibitemShut {NoStop}%
\bibitem [{\citenamefont {Chen}\ \emph {et~al.}(2021)\citenamefont {Chen},
  \citenamefont {Abbasi}, \citenamefont {Joglekar},\ and\ \citenamefont
  {Murch}}]{chen_quantum_2021}%
  \BibitemOpen
  \bibfield  {author} {\bibinfo {author} {\bibfnamefont {W.}~\bibnamefont
  {Chen}}, \bibinfo {author} {\bibfnamefont {M.}~\bibnamefont {Abbasi}},
  \bibinfo {author} {\bibfnamefont {Y.~N.}\ \bibnamefont {Joglekar}}, \ and\
  \bibinfo {author} {\bibfnamefont {K.~W.}\ \bibnamefont {Murch}},\ }\href
  {\doibase 10.1103/PhysRevLett.127.140504} {\bibfield  {journal} {\bibinfo
  {journal} {Phys. Rev. Lett.}\ }\textbf {\bibinfo {volume} {127}},\ \bibinfo
  {pages} {140504} (\bibinfo {year} {2021})}\BibitemShut {NoStop}%
\bibitem [{\citenamefont {Khandelwal}\ \emph {et~al.}(2021)\citenamefont
  {Khandelwal}, \citenamefont {Brunner},\ and\ \citenamefont
  {Haack}}]{khandelwal_signatures_2021}%
  \BibitemOpen
  \bibfield  {author} {\bibinfo {author} {\bibfnamefont {S.}~\bibnamefont
  {Khandelwal}}, \bibinfo {author} {\bibfnamefont {N.}~\bibnamefont {Brunner}},
  \ and\ \bibinfo {author} {\bibfnamefont {G.}~\bibnamefont {Haack}},\ }\href
  {\doibase 10.1103/PRXQuantum.2.040346} {\bibfield  {journal} {\bibinfo
  {journal} {PRX Quantum}\ }\textbf {\bibinfo {volume} {2}},\ \bibinfo {pages}
  {040346} (\bibinfo {year} {2021})}\BibitemShut {NoStop}%
\bibitem [{\citenamefont {Larson}\ and\ \citenamefont
  {Qvarfort}(2023)}]{larson_exceptional_2023}%
  \BibitemOpen
  \bibfield  {author} {\bibinfo {author} {\bibfnamefont {J.}~\bibnamefont
  {Larson}}\ and\ \bibinfo {author} {\bibfnamefont {S.}~\bibnamefont
  {Qvarfort}},\ }\href {\doibase 10.1142/S1230161223500087} {\bibfield
  {journal} {\bibinfo  {journal} {Open Systems \& Information Dynamics}\
  }\textbf {\bibinfo {volume} {30}},\ \bibinfo {pages} {2350008} (\bibinfo
  {year} {2023})}\BibitemShut {NoStop}%
\bibitem [{\citenamefont {Chen}\ \emph {et~al.}(2022)\citenamefont {Chen},
  \citenamefont {Abbasi}, \citenamefont {Ha}, \citenamefont {Erdamar},
  \citenamefont {Joglekar},\ and\ \citenamefont
  {Murch}}]{chen_decoherence-induced_2022}%
  \BibitemOpen
  \bibfield  {author} {\bibinfo {author} {\bibfnamefont {W.}~\bibnamefont
  {Chen}}, \bibinfo {author} {\bibfnamefont {M.}~\bibnamefont {Abbasi}},
  \bibinfo {author} {\bibfnamefont {B.}~\bibnamefont {Ha}}, \bibinfo {author}
  {\bibfnamefont {S.}~\bibnamefont {Erdamar}}, \bibinfo {author} {\bibfnamefont
  {Y.~N.}\ \bibnamefont {Joglekar}}, \ and\ \bibinfo {author} {\bibfnamefont
  {K.~W.}\ \bibnamefont {Murch}},\ }\href {\doibase
  10.1103/PhysRevLett.128.110402} {\bibfield  {journal} {\bibinfo  {journal}
  {Phys. Rev. Lett.}\ }\textbf {\bibinfo {volume} {128}},\ \bibinfo {pages}
  {110402} (\bibinfo {year} {2022})}\BibitemShut {NoStop}%
\bibitem [{\citenamefont {Kumar}\ \emph {et~al.}(2022)\citenamefont {Kumar},
  \citenamefont {Snizhko}, \citenamefont {Gefen},\ and\ \citenamefont
  {Rosenow}}]{kumar_optimized_2022}%
  \BibitemOpen
  \bibfield  {author} {\bibinfo {author} {\bibfnamefont {P.}~\bibnamefont
  {Kumar}}, \bibinfo {author} {\bibfnamefont {K.}~\bibnamefont {Snizhko}},
  \bibinfo {author} {\bibfnamefont {Y.}~\bibnamefont {Gefen}}, \ and\ \bibinfo
  {author} {\bibfnamefont {B.}~\bibnamefont {Rosenow}},\ }\href {\doibase
  10.1103/PhysRevA.105.L010203} {\bibfield  {journal} {\bibinfo  {journal}
  {Phys. Rev. A}\ }\textbf {\bibinfo {volume} {105}},\ \bibinfo {pages}
  {L010203} (\bibinfo {year} {2022})}\BibitemShut {NoStop}%
\bibitem [{\citenamefont {Bu}\ \emph {et~al.}(2023)\citenamefont {Bu},
  \citenamefont {Zhang}, \citenamefont {Ding}, \citenamefont {Li},
  \citenamefont {Zhang}, \citenamefont {Wang}, \citenamefont {Ding},
  \citenamefont {Yuan}, \citenamefont {Chen}, \citenamefont {\"Ozdemir},
  \citenamefont {Zhou}, \citenamefont {Jing},\ and\ \citenamefont
  {Feng}}]{bu_enhancement_2023}%
  \BibitemOpen
  \bibfield  {author} {\bibinfo {author} {\bibfnamefont {J.-T.}\ \bibnamefont
  {Bu}}, \bibinfo {author} {\bibfnamefont {J.-Q.}\ \bibnamefont {Zhang}},
  \bibinfo {author} {\bibfnamefont {G.-Y.}\ \bibnamefont {Ding}}, \bibinfo
  {author} {\bibfnamefont {J.-C.}\ \bibnamefont {Li}}, \bibinfo {author}
  {\bibfnamefont {J.-W.}\ \bibnamefont {Zhang}}, \bibinfo {author}
  {\bibfnamefont {B.}~\bibnamefont {Wang}}, \bibinfo {author} {\bibfnamefont
  {W.-Q.}\ \bibnamefont {Ding}}, \bibinfo {author} {\bibfnamefont {W.-F.}\
  \bibnamefont {Yuan}}, \bibinfo {author} {\bibfnamefont {L.}~\bibnamefont
  {Chen}}, \bibinfo {author} {\bibfnamefont {i.~m. c.~K.}\ \bibnamefont
  {\"Ozdemir}}, \bibinfo {author} {\bibfnamefont {F.}~\bibnamefont {Zhou}},
  \bibinfo {author} {\bibfnamefont {H.}~\bibnamefont {Jing}}, \ and\ \bibinfo
  {author} {\bibfnamefont {M.}~\bibnamefont {Feng}},\ }\href {\doibase
  10.1103/PhysRevLett.130.110402} {\bibfield  {journal} {\bibinfo  {journal}
  {Phys. Rev. Lett.}\ }\textbf {\bibinfo {volume} {130}},\ \bibinfo {pages}
  {110402} (\bibinfo {year} {2023})}\BibitemShut {NoStop}%
\bibitem [{\citenamefont {Mukamel}\ \emph {et~al.}(2023)\citenamefont
  {Mukamel}, \citenamefont {Li},\ and\ \citenamefont
  {Galperin}}]{mukamel_exceptional_2023}%
  \BibitemOpen
  \bibfield  {author} {\bibinfo {author} {\bibfnamefont {S.}~\bibnamefont
  {Mukamel}}, \bibinfo {author} {\bibfnamefont {A.}~\bibnamefont {Li}}, \ and\
  \bibinfo {author} {\bibfnamefont {M.}~\bibnamefont {Galperin}},\ }\href
  {\doibase 10.1063/5.0142022} {\bibfield  {journal} {\bibinfo  {journal} {J.
  Chem. Phys.}\ }\textbf {\bibinfo {volume} {158}},\ \bibinfo {pages} {154106}
  (\bibinfo {year} {2023})},\ \bibinfo {note} {publisher: American Institute of
  Physics}\BibitemShut {NoStop}%
\bibitem [{\citenamefont {Am-Shallem}\ \emph {et~al.}(2015)\citenamefont
  {Am-Shallem}, \citenamefont {Kosloff},\ and\ \citenamefont
  {Moiseyev}}]{am-shallem_exceptional_2015}%
  \BibitemOpen
  \bibfield  {author} {\bibinfo {author} {\bibfnamefont {M.}~\bibnamefont
  {Am-Shallem}}, \bibinfo {author} {\bibfnamefont {R.}~\bibnamefont {Kosloff}},
  \ and\ \bibinfo {author} {\bibfnamefont {N.}~\bibnamefont {Moiseyev}},\
  }\href {http://stacks.iop.org/1367-2630/17/i=11/a=113036} {\bibfield
  {journal} {\bibinfo  {journal} {New J. Phys.}\ }\textbf {\bibinfo {volume}
  {17}},\ \bibinfo {pages} {113036} (\bibinfo {year} {2015})}\BibitemShut
  {NoStop}%
\bibitem [{\citenamefont {Hatano}(2019)}]{hatano_exceptional_2019}%
  \BibitemOpen
  \bibfield  {author} {\bibinfo {author} {\bibfnamefont {N.}~\bibnamefont
  {Hatano}},\ }\href {\doibase 10.1080/00268976.2019.1593535} {\bibfield
  {journal} {\bibinfo  {journal} {Mol. Phys.}\ }\textbf {\bibinfo {volume}
  {117}},\ \bibinfo {pages} {2121} (\bibinfo {year} {2019})}\BibitemShut
  {NoStop}%
\bibitem [{\citenamefont {Perina~Jr}\ \emph {et~al.}(2022)\citenamefont
  {Perina~Jr}, \citenamefont {Miranowicz}, \citenamefont {Chimczak},\ and\
  \citenamefont {Kowalewska-Kudlaszyk}}]{perinajr_quantum_2022}%
  \BibitemOpen
  \bibfield  {author} {\bibinfo {author} {\bibfnamefont {J.}~\bibnamefont
  {Perina~Jr}}, \bibinfo {author} {\bibfnamefont {A.}~\bibnamefont
  {Miranowicz}}, \bibinfo {author} {\bibfnamefont {G.}~\bibnamefont
  {Chimczak}}, \ and\ \bibinfo {author} {\bibfnamefont {A.}~\bibnamefont
  {Kowalewska-Kudlaszyk}},\ }\href {\doibase 10.22331/q-2022-12-22-883}
  {\bibfield  {journal} {\bibinfo  {journal} {{Quantum}}\ }\textbf {\bibinfo
  {volume} {6}},\ \bibinfo {pages} {883} (\bibinfo {year} {2022})}\BibitemShut
  {NoStop}%
\bibitem [{\citenamefont {Tay}(2023)}]{tay_liouvillian_2023}%
  \BibitemOpen
  \bibfield  {author} {\bibinfo {author} {\bibfnamefont {B.~A.}\ \bibnamefont
  {Tay}},\ }\href {\doibase https://doi.org/10.1016/j.physa.2023.128736}
  {\bibfield  {journal} {\bibinfo  {journal} {Physica A}\ }\textbf {\bibinfo
  {volume} {620}},\ \bibinfo {pages} {128736} (\bibinfo {year}
  {2023})}\BibitemShut {NoStop}%
\bibitem [{\citenamefont {Haug}\ and\ \citenamefont
  {Jauho}(2008)}]{haug_quantum_2008}%
  \BibitemOpen
  \bibfield  {author} {\bibinfo {author} {\bibfnamefont {H.}~\bibnamefont
  {Haug}}\ and\ \bibinfo {author} {\bibfnamefont {A.-P.}\ \bibnamefont
  {Jauho}},\ }\href@noop {} {\emph {\bibinfo {title} {Quantum {Kinetics} in
  {Transport} and {Optics} of {Semiconductors}}}},\ \bibinfo {edition} {second,
  substantially revised edition}\ ed.\ (\bibinfo  {publisher} {Springer},\
  \bibinfo {address} {Berlin Heidelberg},\ \bibinfo {year} {2008})\BibitemShut
  {NoStop}%
\bibitem [{\citenamefont {Lipavsk{\' y}}\ \emph {et~al.}(1986)\citenamefont
  {Lipavsk{\' y}}, \citenamefont {{\v S}pi{\v c}ka},\ and\ \citenamefont
  {Velick{\' y}}}]{lipavsk_y_generalized_1986}%
  \BibitemOpen
  \bibfield  {author} {\bibinfo {author} {\bibfnamefont {P.}~\bibnamefont
  {Lipavsk{\' y}}}, \bibinfo {author} {\bibfnamefont {V.}~\bibnamefont {{\v
  S}pi{\v c}ka}}, \ and\ \bibinfo {author} {\bibfnamefont {B.}~\bibnamefont
  {Velick{\' y}}},\ }\href {\doibase 10.1103/PhysRevB.34.6933} {\bibfield
  {journal} {\bibinfo  {journal} {Phys. Rev. B}\ }\textbf {\bibinfo {volume}
  {34}},\ \bibinfo {pages} {6933} (\bibinfo {year} {1986})}\BibitemShut
  {NoStop}%
\bibitem [{\citenamefont {Esposito}\ and\ \citenamefont
  {Galperin}(2009)}]{esposito_transport_2009}%
  \BibitemOpen
  \bibfield  {author} {\bibinfo {author} {\bibfnamefont {M.}~\bibnamefont
  {Esposito}}\ and\ \bibinfo {author} {\bibfnamefont {M.}~\bibnamefont
  {Galperin}},\ }\href {\doibase 10.1103/PhysRevB.79.205303} {\bibfield
  {journal} {\bibinfo  {journal} {Phys. Rev. B}\ }\textbf {\bibinfo {volume}
  {79}},\ \bibinfo {pages} {205303} (\bibinfo {year} {2009})}\BibitemShut
  {NoStop}%
\bibitem [{\citenamefont {Esposito}\ and\ \citenamefont
  {Galperin}(2010)}]{esposito_self-consistent_2010}%
  \BibitemOpen
  \bibfield  {author} {\bibinfo {author} {\bibfnamefont {M.}~\bibnamefont
  {Esposito}}\ and\ \bibinfo {author} {\bibfnamefont {M.}~\bibnamefont
  {Galperin}},\ }\href {\doibase 10.1021/jp103369s} {\bibfield  {journal}
  {\bibinfo  {journal} {J. Phys. Chem. C}\ }\textbf {\bibinfo {volume} {114}},\
  \bibinfo {pages} {20362} (\bibinfo {year} {2010})}\BibitemShut {NoStop}%
\bibitem [{\citenamefont {Fuchs}\ \emph {et~al.}(2014)\citenamefont {Fuchs},
  \citenamefont {Main}, \citenamefont {Cartarius},\ and\ \citenamefont
  {Wunner}}]{fuchs_harmonic_2014}%
  \BibitemOpen
  \bibfield  {author} {\bibinfo {author} {\bibfnamefont {J.}~\bibnamefont
  {Fuchs}}, \bibinfo {author} {\bibfnamefont {J.}~\bibnamefont {Main}},
  \bibinfo {author} {\bibfnamefont {H.}~\bibnamefont {Cartarius}}, \ and\
  \bibinfo {author} {\bibfnamefont {G.}~\bibnamefont {Wunner}},\ }\href@noop {}
  {\bibfield  {journal} {\bibinfo  {journal} {J. Phys. A: Math. and Theor.}\
  }\textbf {\bibinfo {volume} {47}},\ \bibinfo {pages} {125304} (\bibinfo
  {year} {2014})}\BibitemShut {NoStop}%
\bibitem [{\citenamefont {Frigo}\ and\ \citenamefont {Johnson}(2005)}]{FFTW}%
  \BibitemOpen
  \bibfield  {author} {\bibinfo {author} {\bibfnamefont {M.}~\bibnamefont
  {Frigo}}\ and\ \bibinfo {author} {\bibfnamefont {S.~G.}\ \bibnamefont
  {Johnson}},\ }\href@noop {} {\bibfield  {journal} {\bibinfo  {journal} {Proc.
  IEEE}\ }\textbf {\bibinfo {volume} {93}},\ \bibinfo {pages} {216} (\bibinfo
  {year} {2005})}\BibitemShut {NoStop}%
\bibitem [{\citenamefont {Stan}\ \emph {et~al.}(2009)\citenamefont {Stan},
  \citenamefont {Dahlen},\ and\ \citenamefont {van Leeuwen}}]{stan_time_2009}%
  \BibitemOpen
  \bibfield  {author} {\bibinfo {author} {\bibfnamefont {A.}~\bibnamefont
  {Stan}}, \bibinfo {author} {\bibfnamefont {N.~E.}\ \bibnamefont {Dahlen}}, \
  and\ \bibinfo {author} {\bibfnamefont {R.}~\bibnamefont {van Leeuwen}},\
  }\href {\doibase 10.1063/1.3127247} {\bibfield  {journal} {\bibinfo
  {journal} {J. Chem. Phys.}\ }\textbf {\bibinfo {volume} {130}},\ \bibinfo
  {pages} {224101} (\bibinfo {year} {2009})}\BibitemShut {NoStop}%
\bibitem [{\citenamefont {Mandelshtam}\ and\ \citenamefont
  {Taylor}(1997)}]{mandelshtam_harmonic_1997}%
  \BibitemOpen
  \bibfield  {author} {\bibinfo {author} {\bibfnamefont {V.~A.}\ \bibnamefont
  {Mandelshtam}}\ and\ \bibinfo {author} {\bibfnamefont {H.~S.}\ \bibnamefont
  {Taylor}},\ }\href {\doibase 10.1063/1.475324} {\bibfield  {journal}
  {\bibinfo  {journal} {J. Chem. Phys.}\ }\textbf {\bibinfo {volume} {107}},\
  \bibinfo {pages} {6756} (\bibinfo {year} {1997})}\BibitemShut {NoStop}%
\end{thebibliography}

%
\end{document}